**ARTICLE TYPE**

# Investigating four new candidate redback pulsars discovered in the image plane

F. Petrou,[1] N. Hurley-Walker,[1] S. McSweeney,[1] S. Sett,[1] R. Kyer,[2] C. M. Tan,[1] Y. Maan,[3] A. Bahramian,[1] D. Dobie,[4,5] D. L. Kaplan,[6] A. Zic,[7] J. S. Deneva,[8] Tara Murphy,[4,5] E. J. Polisensky,[9] and A. Anumarlapudi[6]

[1]International Centre for Radio Astronomy Research, Curtin University, Bentley, WA, Australia
[2]Center for Data Intensive and Time Domain Astronomy, Department of Physics and Astronomy, Michigan State University, East Lansing, MI 48824, USA
[3]National Centre for Radio Astrophysics, Tata Institute of Fundamental Research, Post Bag 3, Ganeshkhind, Pune - 411007, India
[4]Sydney Institute for Astronomy, School of Physics, The University of Sydney, New South Wales 2006, Australia
[5]ARC Centre of Excellence for Gravitational Wave Discovery (OzGrav), Hawthorn, VIC 3122, Australia
[6]Center for Gravitation, Cosmology and Astrophysics, Department of Physics, University of Wisconsin-Milwaukee, P.O. Box 413, Milwaukee, WI 53201, USA
[7]Australia Telescope National Facility, CSIRO, Space and Astronomy, PO Box 76, Epping, NSW 1710, Australia
[8]George Mason University, resident at Space Science Division, Naval Research Laboratory, Washington, DC 20375, USA
[9]U.S. Naval Research Laboratory, 4555 Overlook Ave. SW, Washington, DC 20375, USA
Author for correspondence: F. Petrou, Email: flora.petrou@postgrad.curtin.edu.au.

## Abstract

This paper reports the discovery and follow-up of four candidate redback spider pulsars: GPM J1723–33, GPM J1734–28, GPM J1752–30 and GPM J1815–14, discovered with the Murchison Widefield Array (MWA) from an imaging survey of the Galactic Plane. These sources are considered to be redback candidates based on their eclipsing variability, steep negative spectral indices, and potential *Fermi* γ-ray associations, with GPM J1723–33 and GPM J1815–14 lying within a *Fermi* 95 % error ellipse. Follow-up pulsation searches with MeerKAT confirmed pulsations from GPM J1723–33, while the non-detections of the other three are likely due to scattering by material ablated from their companion stars. We identify possible orbital periods by applying folding algorithms to the light curves and determine that all sources have short orbital periods (<24 hours), consistent with redback spider systems. Following up on the sources at multiple radio frequencies revealed that the sources exhibit frequency-dependent eclipses, with longer eclipses observed at lower frequencies. We place broad constraints on the eclipse medium, ruling out induced Compton scattering and cyclotron absorption. Three sources are spatially consistent with optical sources in the Dark Energy Camera Plane Survey imaging, which may contain the optical counterparts. Each field is affected by strong dust extinction, and follow-up with large telescopes is needed to identify the true counterparts. Identifying potential radio counterparts to four previously unassociated *Fermi* sources brings us closer to understanding the origin of the unexplained γ-ray excess in the Galactic Centre.

## 1. Introduction

Millisecond pulsars (MSPs) are a population of radio pulsars, typically found in binary systems, that attain their rapid rotational periods ($\leq 30$ ms) through accretion of matter from a companion star through Roche-lobe overflow. This is sometimes referred to as a recycling scenario and was first proposed by Alpar et al. (1982). During this accretion stage, the binary system can be detected as an X-ray binary (XRB). Once accretion ceases, the system transitions to a rotation-powered phase, during which the neutron star behaves like a radio pulsar. Systems that alternate between accretion-powered and rotation-powered states are known as transitional millisecond pulsars (tMSPs) (Papitto and de Martino 2022). These rare objects link XRBs and binary MSPs. Once the companion star completely detaches from its Roche lobe, the system transitions into a rotation-powered MSP. The pulsar wind can now begin to ablate its companion star, gradually stripping away its mass and forming a "spider" binary (Kluzniak et al. 1988; van den Heuvel and van Paradijs 1988). These systems are characterised by their low-mass companion stars and short orbital periods ($P_{\rm orb}$ <24 hours).

Spider systems are comprised of two sub-classes; black widows (BW) with low mass companion stars of <$0.1\,M_\odot$ and redbacks (RB) with companion masses $0.1$–$0.4\,M_\odot$ (Roberts 2013). Currently, only 26 RBs and 44 BWs have been discovered in the Galactic disc and Globular Clusters (Hui and Li 2019; Lee et al. 2023).

Spider systems are challenging to detect in traditional pulsar surveys for several reasons. The primary issue is the large and rapidly changing acceleration due to their short orbital periods, causing the apparent spin frequency to vary significantly throughout the observation. Additionally, eclipses can complicate detection, as typical pulsar surveys have short dwell times, and BWs/RBs are eclipsed for ∼10% and ∼50% of their orbit, respectively (e.g. Fruchter, Stinebring, and Taylor 1988). Lastly, the ionised material around the pulsar can disperse, scatter, or absorb the radio emission, further hindering detection (Roy et al. 2015; Kansabanik et al. 2021).

An in-depth exploration of possible eclipse mechanisms by Thompson et al. 1994 highlights that different systems may have different eclipse mechanisms. Spider systems exhibit frequency dependent eclipses, with eclipses lasting slightly longer at lower frequencies (e.g. Thompson et al. 1994; Main et al. 2018; Lin et al. 2021; Kansabanik et al. 2021). At higher frequencies (up to 4 GHz), only three known spider systems - PSRs J1723–2837, J1731–1847 and J1908+2105 (Crawford et al. 2013; Bates et al. 2011; Ghosh et al. 2025) - are observed to eclipse. The frequency dependence of eclipse duration is a key factor in distinguishing between different eclipse mechanisms, and wide-bandwidth observations are essential for detecting any changes in the spectral features caused by the eclipse material. For instance, Kansabanik et al. (2021) used the upgraded Giant Meterwave Radio Telescope (uGMRT; Swarup 1991) to investigate the eclipse mechanism of PSR J1544+4937 and found that synchrotron absorption by relativistic electrons was the favoured mechanism, as it provided the best fit to the frequency-dependent eclipse spectra and allowed them to estimate a magnetic field strength of 13 G, consistent with the pulsar's environment.

The eclipsing material introduces inhomogeneities in the plasma's electron density, which can cause the radio waves to scatter. This scattering results in the broadening and smearing of the pulsar's radio pulses, making it more difficult to detect the regular, sharp pulses. The extent of the scattering depends on the density of the plasma, with denser regions causing stronger dispersion and signal distortion. However, the pulsars remain observable in the image domain, where eclipse-like variability is still detected (e.g. Zic et al. 2024; Thongmeearkom et al. 2024). Recently, Thongmeearkom et al. (2024) conducted a targeted search of six RB candidates and confirmed pulsations in three of them. One of these, PSR J0838-2827, had previously remained undetected despite repeated observations. The earlier non-detections are thought to be due to longer-duration eclipses, which are likely caused by variations in the eclipsing material. This suggests that the material is highly dynamic, with changes in its density or distribution potentially extending the eclipse duration and hindering pulsar detection. Studying this material provides insight into the interaction between the pulsar's wind and the eclipsing medium, as well as the companion star's mass loss rate.

Traditionally, binary pulsars have been found through untargeted time-domain surveys (e.g. Wang et al. 2025; Padmanabh et al. 2024; Lewis et al. 2023; Lorimer et al. 2006). However, this approach may not be the most efficient for identifying such systems due to their variability and the rapidly changing accelerations induced by their short orbital periods. With the development of wide-bandwidth, low-frequency radio telescopes such as the SKAO telescopes and the precursors, image-domain searches are becoming more promising for detecting eclipsing systems, as these instruments offer increased sensitivity. The spectra of MSPs are typically well described by a power law, with flux density ($S_\nu$) following $S_\nu \propto \nu^\alpha$. The spectral index ($\alpha$) for MSPs is generally around $-1.86(6)$ (Gitika et al. 2023). This steep-spectrum characteristic makes low-frequency telescopes particularly sensitive to detecting such sources.

After the discovery of the first MSP PSR B1937+21 (Backer et al. 1982), it was soon predicted that MSPs could emit $\gamma$-ray emission (Usov 1983). This wasn't observationally confirmed until the launch of the Large Area Telescope (LAT) (Atwood et al. 2009; Abdo et al. 2013) onboard the *Fermi*-gamma ray telescope (*Fermi*), which revealed a $\gamma$-ray excess coming from the centre of our galaxy (Goodenough and Hooper 2009; Ajello et al. 2016). This excess has sparked significant interest, with one possible explanation being a hidden population of MSPs emitting $\gamma$-rays. The *Fermi*-LAT has detected 294 $\gamma$-ray pulsars across the sky, of which 144 are identified as MSPs, 32 as BWs and 13 RBs (Smith et al. 2023). However, there are still over 2000 unassociated Fermi $\gamma$-ray sources whose nature remains uncertain. While most of these are expected to be blazars, some could represent a hidden population of Galactic MSPs. Given that MSPs are part of an old stellar population, it is expected that a hidden concentration of them could reside in the Galactic Bulge. Recent discoveries of MSPs at the locations of some of these unassociated sources (e.g., Thongmeearkom et al. 2024; Himes and Jagannathan 2024; Frail et al. 2018) support the hypothesis that the Galactic Center $\gamma$-ray excess is caused by a population of MSPs. However, an alternative hypothesis for the excess is dark matter annihilation (e.g., Abazajian et al. 2014; Abazajian and Keeley 2016).

In this paper, we report the radio image-plane detection of four candidate RB pulsars GPM J1723–33, GPM J1734–28, GPM J1752–30 and GPM J1815–14. The details of the discovery and follow-up observations are provided in Section 2, with source detections presented in Section 3. We also present the closest unassociated *Fermi* $\gamma$-ray source. The corresponding analysis techniques are described in Section 4, where we present a comprehensive sampling of the spectral energy distribution (SED) for each source, identify possible eclipse mechanisms, and provide initial constraints on the orbital parameters of the systems. We also examine the luminosity distribution of known RBs and compare it to our sample to obtain an estimate of future RB detections with SKA–low. Finally, we summarise in Section 6.

## 2. Observations

This section outlines the discovery of the four sources discussed in this paper, followed by subsections detailing survey and follow-up observations of each source, covering radio imaging and time–domain data. The selected facilities provide a wide sampling of the SED of each source, enabling comparisons with different eclipse models tested in Section 4.3. Observations were carried out using MWA (200 MHz), uGMRT (550–750 MHz), VAST (888 MHz), Parkes (704–4032 MHz), ATCA (1100–3100 MHz) and MeerKAT (2625–3500 MHz). Optical and near-infrared counterparts for the sources were searched for in existing surveys, as detailed in Section 2.7. Table 2 gives all essential details for the observations.

### 2.1 MWA

The Murchison Widefield Array (Tingay et al. 2013; Wayth et al. 2018, MWA;) is a low-frequency (80––300 MHz) precursor to the SKAO located at the Inyarrimanha Ilgari Bundara, the CSIRO's Murchison Radio-astronomy Observatory, in Western Australia. One project carried out with the MWA is the Galactic Plane Monitor (GPM). These observations were taken between June–September 2022 and June–September 2024 under project code G0080 (see Methods of Hurley-Walker et al. 2023). The prime objective of the campaign was to search for long-period transients by scanning the Galactic plane over $285° < l < 65°$, $|b| < ±15°$ on a bi-weekly cadence at 185–215 MHz; a full description will be released by Hurley-Walker et al., in prep. Observations were undertaken in five-minute "snapshots" and mosaics were formed for each night of observing. For any given region, the effective integration time was 30 min for the data taken in 2022, and 45 min in 2024. The four sources discussed in this paper were discovered in these data as clearly varying between "on" and "off" states; the survey and selection process details will be presented in a separate publication (Hurley-Walker et al., in prep).

## 2.2 uGMRT

The upgraded Giant Meterwave Radio Telescope (uGMRT; Swarup 1991; Gupta et al. 2017) comprises of 30 antennas with a 45-m diameter, located in Pune, India. Our observations were taken between October 2023 and December 2024, using band-4 (550–750 MHz). We aimed to schedule the observations just before or after the candidate pulsars reach superior conjunction relative to their companions (i.e. as the candidate pulsar enters or exits eclipse), so that we can study the eclipsing material. Each observation commenced with a ~5 minute scan of a bright, compact radio source (e.g. 3C286) for flux and bandpass calibration. The targets were observed for ~30 minutes each, bookended with a ~5 minute nearby phase calibrator scan. Calibration and flagging were done using Common Astronomy Software Applications (CASA; CASA Team et al. 2022). Imaging was done using WSClean (Offringa et al. 2014), performing multi-frequency synthesis imaging, with a Briggs –0.5 weighting scheme (Briggs 1995) and joint-channel deconvolution. The bandwidth was split between 2–10 channels, depending on the image's signal-to-noise ratio (S/N). Aegean[a] (Hancock et al. 2012; Hancock, Trott, and Hurley-Walker 2018) was used to extract the flux densities and associated errors. When the source is eclipsed, the flux density is reported as an upper limit by measuring the pixel value at the location of the source.

Pulsar search mode data were also recorded simultaneously with the interferometric visibility data, using the phased-array observing mode of uGMRT. These data utilised 4096 channels sampling 200 MHz bandwidth centred at 650 MHz, and a sampling time of 81.96 µs. RFIClean (Maan, van Leeuwen, and Vohl 2021) was used for the first level of RFI excision, as well as conversion of the data to SIGPROC filterbank format. A periodicity search for pulsed signals was then conducted using PRESTO (Ransom 2011) on the data recorded in October and November 2023; see Section 3.2 for more details.

## 2.3 ASKAP

The Australian Square Kilometre Array Pathfinder (ASKAP) (Hotan et al. 2021) Variables and Slow Transients Survey (VAST) (Murphy et al. 2021), is a radio survey designed to detect transient sources in the image domain. The Galactic components of VAST observes 42 fields over $|b| < 6°$ and $|l| < 10°$ totalling 1260 deg$^2$. It operates at a central frequency 888 MHz, taking observations roughly every two weeks since November 2022. The integration time for each observation is ~12 minutes, with a typical sensitivity of 0.24 mJybeam$^{-1}$. Science-ready data products are made available through CSIRO ASKAP Science Data Archive (CASDA) (Huynh et al. 2020) and are produced using the ASKAPsoft pipeline (Guzman et al. 2019).

## 2.4 Parkes Murriyang

Murriyang, the Parkes 64-m radio telescope, is a 64-m single-dish radio telescope situated in New South Wales. Our observations utilised the MEDUSA back-end and were taken during available Director's Time (under project code PX095) in the Ultra Wide Bandwidth (UWL) receiver (Hobbs et al. 2020), which offers a continuous frequency range of 704–4032 MHz. The observations were recorded at 64-µs time resolution, with 128 channels per subband and 26 subbands across the UWL band. Each source was observed for approximately 15 minutes. The data was processed using standard processing software, PRESTO performing a periodicity and acceleration search (that is described in detail in Section 3.2).

## 2.5 ATCA

The Australia Telescope Compact Array (ATCA; Wilson et al. 2011), consists of six 22-m antennas located in Narrabri, New South Wales. We took observations using the 16-cm receiver, utilising the Compact Array Broadband Backend (CABB) in the L-band (1100–3100 MHz). The observations were recorded with 1 MHz channels and a correlator integration time of 10 s, under the project code CX544. Each observation observed PKS B1934-638 for ~10 min for flux density and bandpass calibration. A ~2-min scan of a nearby phase calibrator was taken between each ~15-min source scan. The data was flagged and calibrated using standard procedures with CASA and imaged using the task tclean, selecting a Briggs weighting and robust parameter –0.5.

## 2.6 MeerKAT

MeerKAT S-band observations were undertaken under proposal code SCI-20241101-NH-01, using the S4 window at 2625 – 3500 MHz to minimise the effects of scattering. At the time of observing, this sub-band had by far the cleanest RFI environment [b], ensuring that less data is lost to flagging. The observations included typical bandpass, polarisation, and phase calibrators, as well as phase-up and test pulsar observations. Each target source was observed for two 30-minute integrations; we attempted to schedule the observations such that the sources would be observed as close to the candidate pulsars' inferior conjunctions as possible (i.e., least likely to be in eclipse). This was done to increase the likelihood of detecting pulsations. As well as correlator observations undertaken at 8-s/854.492-kHz resolution, we also employed the Pulsar Timing User Supplied Equipment (PTUSE; Bailes et al. 2020) in search mode, using 37.45-μs sampling.

---

[a]. https://github.com/PaulHancock/Aegean

[b]. https://skaafrica.atlassian.net/wiki/spaces/ESDKB/pages/305332225/Radio+Frequency band-statistics

The correlator data were calibrated using the standard SARAO SDP calibration pipeline, and imaged using WSClean. We imaged each 30-min scan separately to yield good signal-to-noise for source position measurements, while also yielding some information about the time-variability of the sources at S-band. A periodicity search for pulsation signal was conducted on the PTUSE data using the PRESTO software (see Section 3.2 for details).

### 2.7 Optical and Near-Infrared

We searched for optical and near-IR counterparts to the four radio positions among existing survey data. The deepest optical imaging of the fields around the first three sources is available through the Dark Energy Camera Plane Survey (DECaPS2; Saydjari et al. 2023), which used 30 s exposures with the 4.0-m Blanco telescope in Chile in 2016 and 2019. As shown in the coadded fields in Figure 1, there is one potential counterpart consistent with the radio position of GPM J1723–33. We identified multiple potential optical counterparts from the DECaPS imaging that are compatible with the radio positions of GPM J1734-28 and GPM J1752-30. We report the DECaPS2 band-merged catalogue measurements for these optical sources in Section 3.4. However, due to significant interstellar extinction in the direction of the sources ($A_z \sim 1.7$ and $2.3$ mag respectively for a distance of 8 kpc, Zucker et al. 2025), it is possible that the true optical counterparts to the radio sources are obscured.

We checked for potential near-IR counterparts among the VISTA Variables in the Via Lactea (VVV) DR5 catalogue and imaging (Minniti et al. 2010), which was performed between 2010-2015 and targeted central parts of the Galaxy primarily in the $K_s$ band with the 4.1-m VISTA telescope in Chile. No catalogue sources are detected within 1.5 arcsec of these positions. Although extinction is less of an obstacle at longer wavelengths, the exposure time used by the VVV survey imaging is quite short ($\sim 4$ sec). Visual inspection of the fields revealed only marginal evidence for flux at the locations of the optical sources in the DeCAPS imaging.

The field of GPM J1815–14 is not covered by the DECaPS2 footprint. To search for nearby optical sources, we inspected deep stack images from Pan-STARRS1, which used a 1.8-m telescope in Hawai'i (Chambers et al. 2016; Flewelling et al. 2020). In the fourth panel of Figure 1, we show a $z$ band stack of 30 and 60 s exposures with a total exposure time of 780 s. The radio position of GPM J1815–14 is close to a bright $z \sim 14.3$ mag foreground star. We note that there is one detection in the Pan-STARRS1 catalog near this radio position in a single epoch, in the $g$ band with $g_{PSF} = 21.7 \pm 0.2$ (were $g_{PSF}$ denotes the Point Spread Function magnitude) offset from the radio position by 0.8 arcsec observed in June 2012. The quality factor for the detection is 0.84, indicating that the detection may be suspect. No other optical sources compatible with the position are detected. The extinction in this direction is extremely high, and so the optical counterpart is likely entirely obscured by dust ($A_z \sim 4.8$ mag at 8 kpc, Green 2018; Green et al. 2019). In the near-IR, the field of GPM J1815–14 was covered by the VVV extended survey in 2016-2019 (VVVX; Saito et al. 2024), and no sources are detected in the catalog or upon inspection of the imaging down to a mean sensitivity of $\sim 18.5$ mag in the $K_s$ band.

### 3. Source Detections

In this section, we report the detections obtained from the observations detailed in Section 2, as well as possible *Fermi* associations.

### 3.1 Fermi gamma-ray Associations

As pulsars are known to emit γ–ray emission, we cross-match our sources with the *Fermi* Fourth Full Catalogue of LAT Sources (4FGL-DR4; Abdollahi et al. 2022). The nearest *Fermi*-4FGL sources found, and corresponding angular separations for our sources, are provided in Table 4 and Figure 2.

Possible counterparts for GPM J1723–33 and GPM J1815–14 lie within the *Fermi* 95 % error ellipse. The nearest 4FGL sources to GPM J1734–28 and GPM J1752–30 lie outside their respective ellipses, at 5 σ and 3 σ respectively.

While this may indicate a low probability of association for the last two sources, Abdollahi et al. (2022) notes that γ-ray source localisation in the Galactic Plane is challenging due to high background emission. To assess the probability that GPM J1734–28 and GPM J1752–30 are associated with their nearest 4FGL sources even though they lie outside the 95% confidence ellipse, we calculate the fractional sky area covered by *Fermi* error ellipses at 5 σ and 3 σ respectively. We consider a 100 deg$^2$ region around our source. The percentage of sky occupied by the ellipses for GPM J1734–28 and GPM J1752–30 is found to be 13 % and 5 % respectively. The results indicate that the likelihood of our source falling that close to a *Fermi* ellipse by chance is relatively low. Therefore, we still consider these as potential counterparts.

The implications of these results are discussed in detail in Section 4.7.

### 3.2 Pulsation Searches

Pulsation searches were carried out with three radio telescopes: uGMRT (550–750 MHz), Parkes (704–4032 MHz) and MeerKAT (2625–3500 MHz), covering a broad frequency range. A periodicity search was performed on the PTUSE data using the PRESTO software suite on the Setonix cluster hosted by Pawsey. Following radio-frequency interference (RFI) mitigation, the data were dedispersed over trial dispersion measures (DMs) from 0–5000 pc cm$^{-3}$. Each DM trial was searched for periodic signals using Fourier-based acceleration searches with harmonic summing up to 8 harmonics and a maximum acceleration parameter of $z_{max} = 300$. Harmonic summing up to 8 was chosen for the acceleration search, as it is generally sufficient for detecting MSPs, particularly in systems such as spiders, which are known to be MSPs and typically do not require higher harmonic summing due to their narrow pulse profiles.

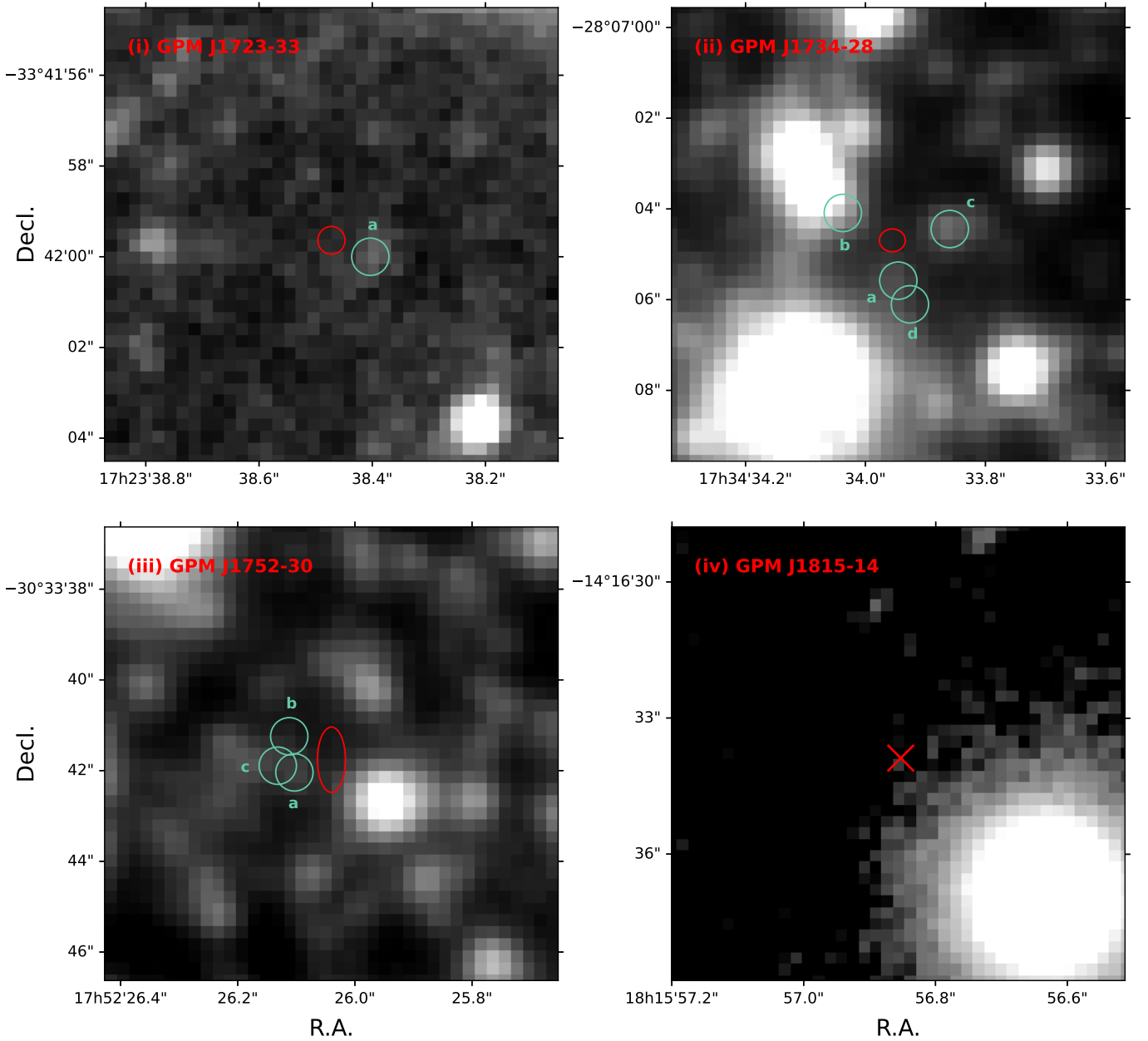

**Figure 1.** Panels (i),(ii),(iii): DECaPS $z$-band images of the fields centred on the radio positions of GPM J1723–33, GPM J1734-28 and GPM J1752–30. The seeing in each image is about 1 arcsec. The radio positions are marked with a red ellipse corresponding to their positional uncertainties, except in the case of GPM J1815–14 where the positional uncertainty is smaller than one pixel in the image. Optical sources in the DECaPS2 band-merged catalogue within 1.5 arcsec of the radio positions are marked by blue circles to guide the eye. The letter appearing next to each optical position corresponds to the catalogue entry in Table 3. Panel (iv): Pan-STARRS1 $z$ band deep stacked image of the field centred on GPM J1815–14. The radio position is marked by a red cross, as its positional uncertainty is the size of about one pixel in this image. Each cutout is $10 \times 10$ arcsec. One faint potential optical counterpart is detected for GPM J1723–33, multiple are detected for GPM J1734–28 and GPM J1752–30, while none are detected for GPM J1815–14.

**Table 1.** The table presents the positional coordinates of the sources discussed in this paper (see Section 3.3), along with the 200 MHz flux density at the pulsar's inferior conjunction, obtained from model fitting of Equation 3. It also includes the orbital period $P_{\rm orb}$, and the reference MJD $T_0$ (see Section 4.2).

| Name | RA J2000 | Dec J2000 | RA error " | Dec error " | $S_{200\,\rm MHz}$ mJy | $P_{\rm orb}$ hr | T0 MJD |
|---|---|---|---|---|---|---|---|
| GPM J1723–33 | 17:23:38.48 | −33:41:59.56 | 0.51 | 0.45 | 58(2) | 5.1477(2) | 59789.6(3) |
| GPM J1734–28 | 17:34:33.96 | −28:07:04.68 | 0.34 | 0.26 | 43(2) | 20.112(2) | 59796.06(4) |
| GPM J1752–30 | 17:52:26.05 | −30:33:41.79 | 0.40 | 0.57 | 46(3) | 17.858(1) | 59807.0(3) |
| GPM J1815–14 | 18:15:56.85 | −14:16:33.9 | 0.11 | 0.10 | 121(3) | 9.81969(2) | 59740.05(7) |

**Table 2.** Summary of the follow-up observations outlined in Section 2.

| Telescope | Project Code | Start Date | $\nu$ (MHz) | Mode |
|---|---|---|---|---|
| MWA | G0008 | 27-Oct-2022 | 200 | Imaging |
| uGMRT | 45_027 and 46_056 | 29-Oct-2023 and 17-Jun-2024 | Band 4 (550–750) | Imaging + Pulsar Search |
| VAST | AS207 | 14-Nov-2023 | 888 | Imaging |
| Parkes | PX095 | 23-Nov-2022 | UWL (704–4032) | Pulsar Search |
| ATCA | CX544 | 10-Jan-2023 | L-band (1100–3100) | Imaging |
| MeerKAT | SCI-20241101-NH-01 | 15-Dec-2024 | S–band (2625–3500) | Imaging + Pulsar Search |

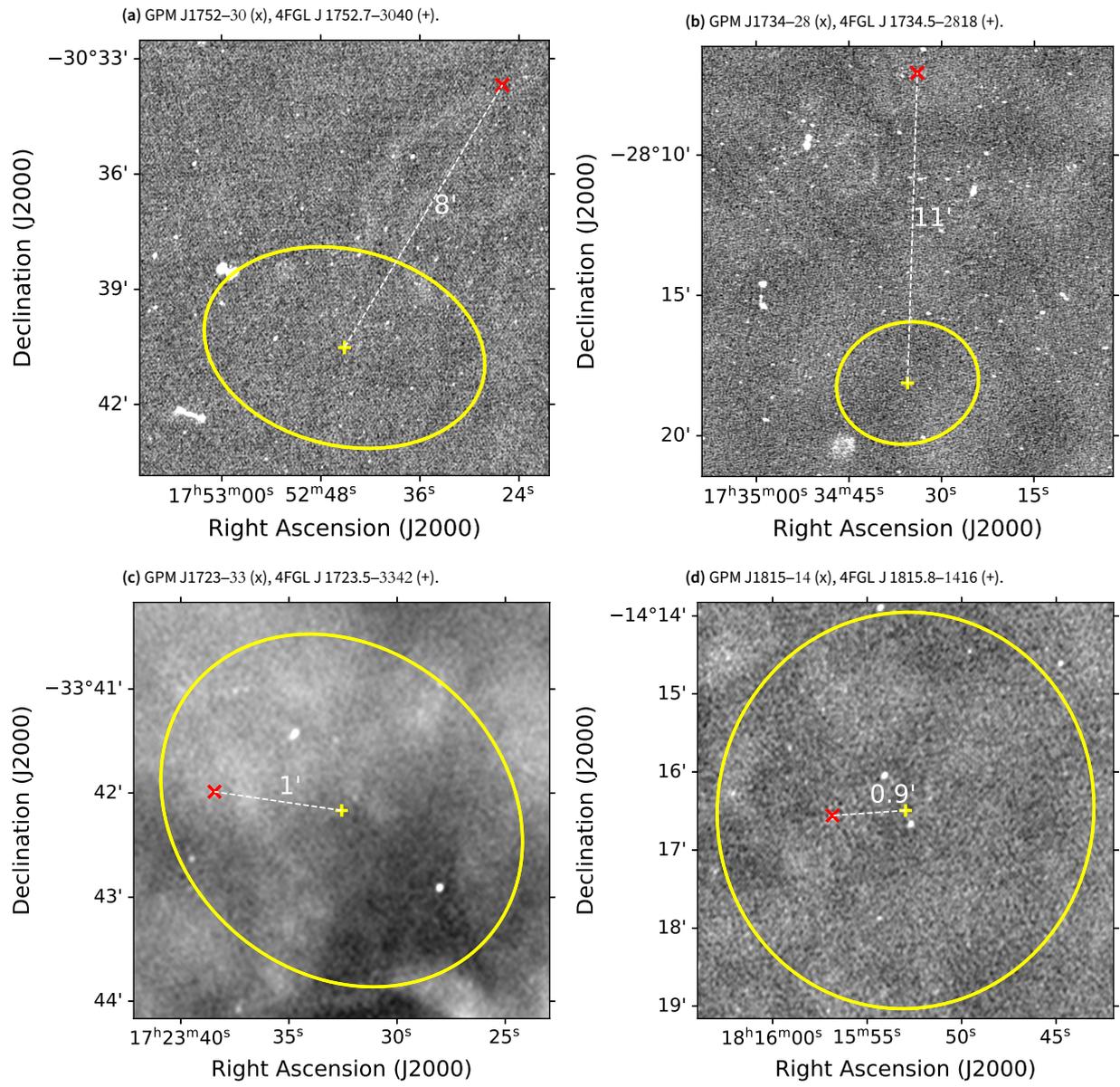

**Figure 2.** MeerKAT S-band images for each source, overlaid with the *Fermi* 95 % error ellipse (shown in yellow) of the closest unassociated $\gamma$-ray source. The angular separation between the radio and $\gamma$-ray sources is indicated. See Table 3.1 for more details.

The acceleration parameter $z$ denotes the maximum number of Fourier frequency bins a pulsar signal can drift due to a constant line-of-sight acceleration during the observation. It is given by:

$$z = \left(\frac{Gm_c^3}{m_{\text{tot}}^2}\right)^{1/3} \left(\frac{2\pi}{P_{\text{orb}}}\right)^{4/3} \quad (1)$$

where $m_c$ is the mass of the companion, $m_{\text{tot}}$ is the total mass of the system and $P_{\text{orb}}$ is the orbital period of the system. The maximum detectable constant line-of-sight acceleration is given by

$$a = \frac{z_{\max} \, cP}{T_{\text{obs}}} \quad (2)$$

where $c$ is the speed of light, $P$ is the pulsar spin period, and $T_{\text{obs}}$ is the observation duration.

We arbitrarily choose a sufficiently high $z_{\max}$ value of 300 for this search, as the expected $z$ for spider pulsars is much smaller. For a example, a RB system with $m_c$ = 0.2 $M_\odot$ and $m_p$ = 1.4 $M_\odot$, and a 12 hr orbital period, has $z \approx 5.7$, corresponding to an acceleration of 16 ms$^{-2}$.

The PTUSE data were then folded to the period and DM of the candidates found in the search, and the diagnostic plots produced were manually inspected.

We also divided the Parkes UWL band into three segments (704–1344, 1344–2368, 2368–4032 MHz) and searched each independently. High-frequency searches help reduce the potential for pulse smearing due to scattering, as the scattering timescale scales with frequency as $\tau_s \propto \nu^{-4}$, while low-frequency searches are particularly effective for detecting pulsars due to their steep negative spectral indices. All candidates were manually inspected in the uGMRT and Parkes data, and no significant pulsar signals were identified.

An $8\sigma$ detection of GPM J1723–33 was made using MeerKAT (see Figure 3). The spin period is 2.11030426(16) ms and the DM is 469(2) pc cm$^{-3}$. The acceleration search measured $z$ = 11, corresponding to a line-of-sight acceleration of 2.1493840(2) ms$^{-2}$ (for $T_{\text{obs}}$ = 1800 s).

All candidates from the MeerKAT observations were manually inspected; however, no convincing pulsar signals were detected. Potential reasons for the non-detection of radio pulses are discussed in Section 5.2.

### 3.3 Radio Imaging

Despite the non-detection of pulsations, we can still obtain valuable insight into these systems through image-domain observations. All four sources were detected as variable sources in the image-domain by multiple radio surveys (MWA, VAST, uGMRT, ATCA, and MeerKAT). The variability observed in the time-integrated snapshot surveys such as VAST and MWA is typically characterised by a bimodal switch between "on" and "off" (see Figure 4 for an example).

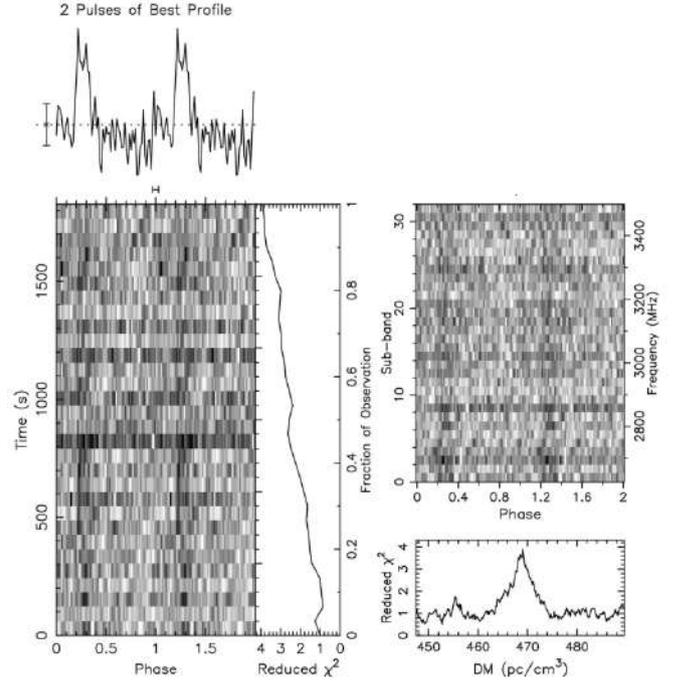

**Figure 3.** Pulsation detection of the MSP GPM J1723–33 with MeerKAT radio telescope. *left:* The main panel shows the evolution of pulsations over time, folded on its spin period of 2.11030426(16) ms. The top subpanel displays the integrated pulse profile, frequency-scrunched to enhance gaussian significance, while the side panel shows the reduced $\chi^2$, indicating the significance of the detection. *right:* The upper plot shows pulse phase as a function of frequency, confirming the broadband nature of the signal. The bottom plot displays the dispersion measure (DM) against $\chi^2$, highlighting the optimal DM solution. This figure is derived from the analysis described in Section 3.2.

The ASKAP positions of the sources were used for initial follow-up observations, determined using weighted averages with an accuracy of $\sim$ 1 arcsec. We refined the positions using MeerKAT observations, which offer improved angular resolution due to their longer baselines and higher sensitivity. We measured the positions using AEGEAN, achieving sub-arcsecond precision. The final positions and associated uncertainties are listed in Table 1. We adopted the MeerKAT positions for our analysis due to their improved accuracy and resolution.

The MeerKAT 30-minute integration images had RMS noise levels between 8–9 µJy beam$^{-1}$. GPM J1723–33 was detected at a flux density of 0.11(2) mJy in one scan, with a non-detection in the other. GPM J1734–28 was detected with a flux density of 60(9) µJy in both scans. GPM J1752–30 was not detected in the first scan but was detected with a flux density of 40(8) µJy in the second. GPM J1815–14 was measured with a flux density is 0.14(1) mJy in both scans.

The uGMRT observations were scheduled to coincide with the time of superior conjunction and successfully captured data during this phase for GPM J1734–28, GPM J1752–30 and GPM J1815–14. The wide-bandwidth of uGMRT enables the data to be segmented by frequency, allowing an investigation of frequency-dependent effects.

We aimed to observe the sources with ATCA at higher radio frequencies and at orbital phases where the candidate pul-sars are in inferior conjunction to their companion. GPM J1723–33 was detected at a flux density of 0.35(6) mJy and GPM J1752–30 at 0.48(6) mJy. However, the S/N ratios (∼6 and ∼8, respectively) were too low to allow for time or frequency segmentation of the data for orbital period analysis. ATCA did not detect GPM J1734–28 and GPM J1815–14 due to noise levels, where the background root-mean-square (RMS) is 30 μJy beam$^{-1}$ and 70 μJy beam$^{-1}$, respectively.

### 3.4 Search for Optical and Near-IR Counterparts

Optical observations of RBs are used to characterise the companion and the binary orbital configuration through detailed light curve modelling and radial velocity studies. We searched for optical and near-IR counterparts to the four new RB candidates in survey imaging and identified potential counterparts to three shown in Figure 1.

Only one faint source in the band-merged DECaPS2 catalogue is detected within 1.5 arcsec of GPM J1723-33. For GPM J1734-28, there are four faint optical sources potentially consistent with the radio position, two of which are visually detected in Figure 1. For GPM J1752–30, there are three faint optical sources consistent with the radio position, two of which are marginally visible upon inspection of the DECaPS images. We summarise the DECaPS2 catalogue measurements of these optical sources in Table 3. No near-IR sources within 1.5 arcsec of the radio positions are detected in the VVV DR5 catalog. Visual inspection of the stacked and epoch imaging showed only marginal evidence for excess brightness at the positions of the DECaPS2 sources, and we found no evidence for variability.

Ground-based follow-up of these fields to determine the correct optical counterparts is challenging due to the high extinction and intrinsic faintness of the targets. Optical observations with larger telescopes or deeper observations in the near-IR bands are needed to detect the sources and variability signatures, which would solidify any association with the radio sources and allow detailed modelling of the binary configurations. With the initial orbital periods determined in Section 4.2, observations could be strategically planned to sample orbital modulation.

The search for optical counterparts to GPM J1815–14 did not yield any optical sources consistent with its radio position. This field is strongly affected by dust extinction, and it is unlikely that the optical counterpart will ever be accessible. Future deep observations in the near-IR may reveal a counterpart.

## 4. Analysis

In this section, we use the source detections detailed in Section 3 to investigate the various properties of the systems, exploring their orbital dynamics, spectral behaviour, potential *Fermi* associations, and absorption characteristics.

### 4.1 Dynamic evolution

Figure 5 shows the MWA GPM data for GPM J1815–14, revealing that the system exhibits significant variability over time. Notably, the source is not detected for most of the 2024 data. RB systems are known to evolve dynamically, as the companion star evaporates due to intense radiation from the pulsar. As the density of the ablated material varies over month-timescales, the amount of flux density that is absorbed or scattered may change, resulting in variations in the observed eclipse duration. This likely explains the decreasing detectability of GPM J1815–14 over time. Therefore, systems like this one may experience varying detectability over time, making consistent detection challenging.

We also considered refractive interstellar scintillation (RISS) as a possible explanation for the observed variability. Hancock et al. (2019) created an all-sky model for RISS, which can predict variability on a variety of timescales, survey locations, and observing frequencies. We use the available code RISS19[c] to obtain an order-of-magnitude estimate for the RISS at the location of GPM J1815–14. We find the scintillation strength is very high near the galactic plane, but the expected flux density modulation at 200 MHz is only a few per cent, with a characteristic timescale of hundreds of years. This suggests that RISS is unlikely to produce significant short-term variability in our source. Intrinsic pulsar variability is another potential cause, but the observed flux density appears correlated with the orbital phase, suggesting an eclipse-related origin rather than random pulsar behaviour.

### 4.2 Orbital Period Analysis

We searched for periodic variability in the MWA and VAST datasets independently for each source, as spider systems often exhibit frequency-dependent eclipses. Using the MJD, flux density, and corresponding flux density errors, we applied the Lafler-Kinman's String Length (LKSL) method (Clarke 2002), using the P4J[d] package. LKSL is well-suited for identifying periodicities in sparse, unevenly sampled light curves. It works by folding the data on a range of trial periods and calculating the string length at each trial; the minima in string length correspond to the best trial period. The P4J implementation inverts this measure so that peaks in the periodogram (LKSL power versus period) correspond to best candidate periods.

We searched trial periods between 0.5 and 24 hours, corresponding to a frequency range of approximately 1 to 0.04167 cycles/day. The frequency grid was sampled at a resolution of $10^{-4}$, and the 12 most significant local peaks were selected. We independently searched for periodicity in the MWA and VAST data, and then compared candidate periods to identify consistencies.

Using this subset of best candidate periods, we phase-folded the flux density measurements from both the VAST and MWA datasets, incorporating the MeerKAT detection and non-detection times, as well as the uGMRT observation window, to assess whether these observations aligned consistently in phase space. We determined the best candidate orbital period as the one most compatible with all the data.

---

c. https://github.com/PaulHancock/RISS19
d. https://github.com/phuijse/P4J

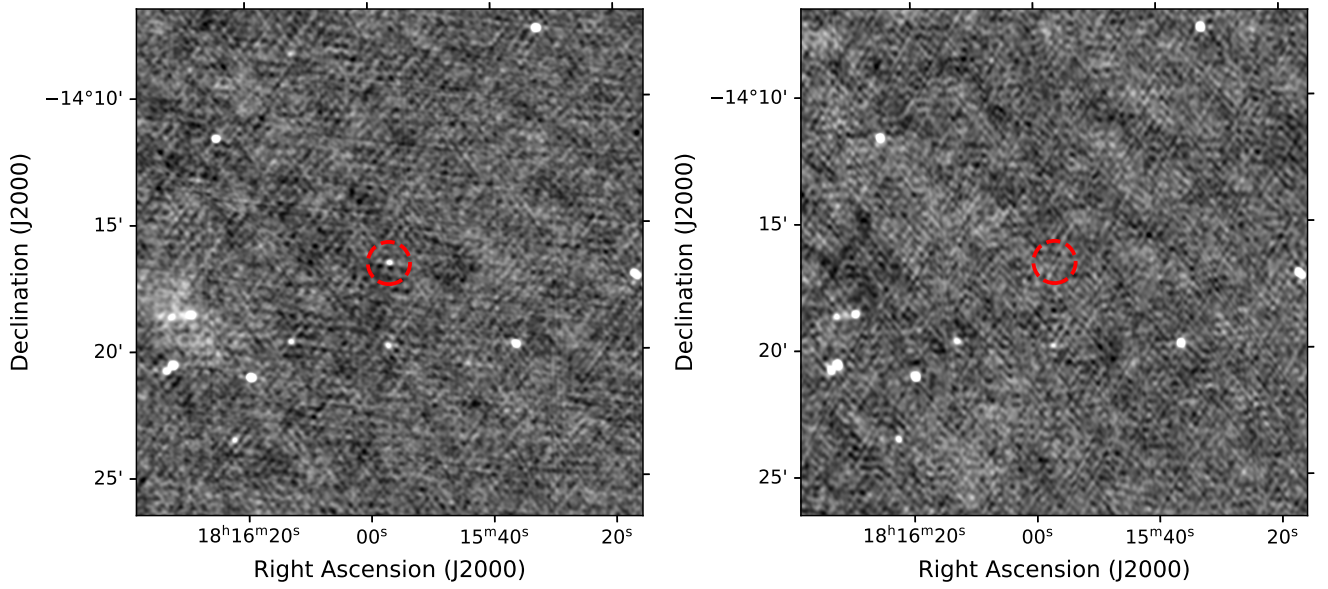

**Figure 4.** *left:* VAST detection of GPM J1815–14 on 2022-12-21T03:20:25.5 UTC. *right:* VAST non-detection of GPM J1815–14 on 2022-11-14T06:57:53.7 UTC.

**Table 3.** Optical sources in the DECaPS2 band-merged catalogue that are compatible with the radio positions of GPM J1723-33, GPM J1734-28 and GPM J1752–30. The Label column indicates the marker used for the optical sources in Figure 1. Unique object identifiers, offsets from the radio positions, and mean $grizY$ background-corrected magnitudes from DECaPS2 are given. All catalogue entries here are > $5\sigma$ detections.

| Name | Label | DECaPS2 obj_id | Offset " | $g$ | $r$ | $i$ | $z$ | $Y$ |
|---|---|---|---|---|---|---|---|---|
| GPM J1723-33 | a | 4522440858633389827 | 0.93 | – | – | – | 21.76 | – |
| GPM J1734-28 | a | 4954927160361310917 | 0.90 | – | 22.45 | 21.57 | 20.63 | – |
|  | b | 4954927160407567054 | 1.25 | – | – | 20.58 | – | – |
|  | c | 4954927160361131656 | 1.29 | 23.46 | 21.40 | 20.46 | 19.62 | 19.44 |
|  | d | 4954927160374882674 | 1.46 | – | 22.79 | 21.72 | – | – |
| GPM J1752-30 | a | 5044647309198495279 | 0.86 | – | – | 20.61 | 20.01 | 19.71 |
|  | b | 5044647309200681966 | 1.06 | – | – | – | – | 20.88 |
|  | c | 5044647309177611897 | 1.19 | – | 21.47 | 20.45 | 19.79 | 19.40 |

**Table 4.** Details of the potential *Fermi* sources associated with each source in this paper. The table lists the separation between the radio and *Fermi* 4FGL sources, along with the semi-major and semi-minor axes of the 95 % error ellipses for the 4FGL sources supplied by *Fermi*.

| Name | $\gamma$-ray | RA J2000 | Dec J2000 | Sep. ′ | Semi Maj. ′ | Semi Min. ′ |
|---|---|---|---|---|---|---|
| GPM J1723–33 | 4FGL J 1723.5–3342c | 17:23:32.59 | −33:42:10.44 | 1 | 4 | 3 |
| GPM J1734–28 | 4FGL J 1734.5–2818 | 17:34:35.52 | −28:18:7.92 | 11 | 5 | 4 |
| GPM J1752–30 | 4FGL J 1752.7–3040 | 17:52:45.17 | −30:40:31.80 | 8 | 7 | 5 |
| GPM J1815–14 | 4FGL J 1815.8–1416 | 18:15:52.99 | −14:16:23.00 | 1 | 5 | 4 |

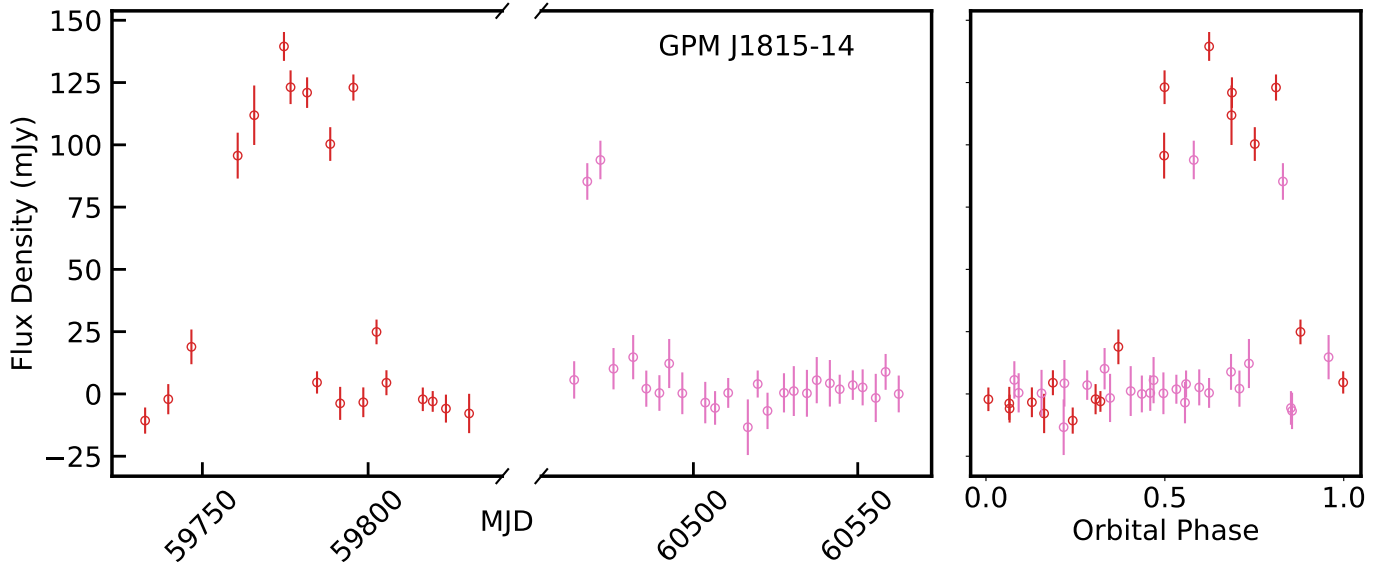

**Figure 5.** *left:* Light curve of GPM J1815–14 using the MWA GPM data at 200 MHz. *right:* The corresponding folded light curve on the orbital period, $P_{\rm orbit}$ =9.81969(2) hrs. The plots demonstrate the increasing eclipse duration over time; see Section 4.1 for details.

Spurious peaks in the periodogram can arise from measurement errors, noise, or random fluctuations in the data. To assess the significance of our candidate period, we applied a bootstrap-based approach. We generated 200 surrogate (bootstrap) light curves by randomly resampling the original data, then computed a periodogram for each, and extracted the 20 highest maxima. An extreme value probability density function (PDF) was then fitted to the distribution of these maxima. From the fitted PDF, we derived 90 and 99 % confidence thresholds, which are shown in the periodograms in Figure 15. The candidate period peaks for GPM J1734–28, GPM J1752–30 and GPM J1815–14 exceed the 90 % threshold, allowing us to confidently conclude that the periodicities are genuine.

Figure 6 presents the folded light curves on the best candidate orbital period found using the LKSL method. All systems are found to have short orbital periods (<1 day) and exhibit eclipsing behaviour. The eclipsing variability is not clearly visible in the VAST GPM J1723–33 data due to large errors caused by deconvolution issues. The final orbital periods and associated uncertainties, derived from the full width at half maximum of the periodogram peak, are given in Table 1.

### 4.3 Eclipse Properties

The properties of the eclipse can be quantified by fitting a double Fermi-Dirac function (see Zic et al. (2024)):

$$S(\phi) = S_0 \left[ \left( \frac{1}{1 + e^{\frac{\phi - \phi_i}{w_i}}} - \frac{1}{1 + e^{\frac{\phi - \phi_e}{w_e}}} \right) \right] \quad (3)$$

where $\phi$ is the phase, $\phi_i$ and $\phi_e$ are the phases at the eclipse ingress and egress, $w_i$ and $w_e$ are the widths of the ingress and egress, and $S_0$ is the flux density at the pulsar's inferior conjunction (i.e., out of eclipse).

For the uGMRT data, and for the VAST data of GPM J1734–28 and GPM J1752–30, which sparsely sampled the orbital phase, we fitted a modified version of Equation 3 to model only the ingress of the eclipse:

$$S(\phi) = S_0 \left[ 1 + \left( \frac{1}{1 + e^{\frac{\phi - \phi_i}{w_i}}} \right) \right] \quad (4)$$

Equations 3 and 4 were fitted to the folded light curves from the MWA, VAST, and uGMRT observations (see Figure 7) using non-linear least-squares fitting via scipy.curve_fit, with bounds and initial guesses chosen to ensure numerical stability. The flux densities are normalised by the maximum flux density obtained from the fit, denoted $S_0$ in Equation 3 and 4. The folded light curves are scaled to a reference MJD, $T_0$ (given in Table 1), corresponding to the time when the candidate pulsars are at superior conjunction relative to their companions. The uncertainty in $T_0$ arises from the propagation of errors from the ingress and egress phase measurements, derived from the covariance matrix of the fitted models.

### 4.4 Radio Spectra

Figure 8 shows the spectra of the sources when they are at inferior conjunction with their companions. For MWA, VAST, and uGMRT, the flux density at inferior conjunction is taken as $S_0$, which is derived from the fitted functions (see Section 4.3). The values for $S_0$ at 200 MHz are provided in Table 1, with the uncertainty on each value assumed to be 20%, accounting for typical calibration and flux scale errors. The data points for ATCA and MeerKAT are taken from the measured flux density at the sources' inferior conjunction.

The spectra were fitted with the following models, taken from Swainston et al. (2022):

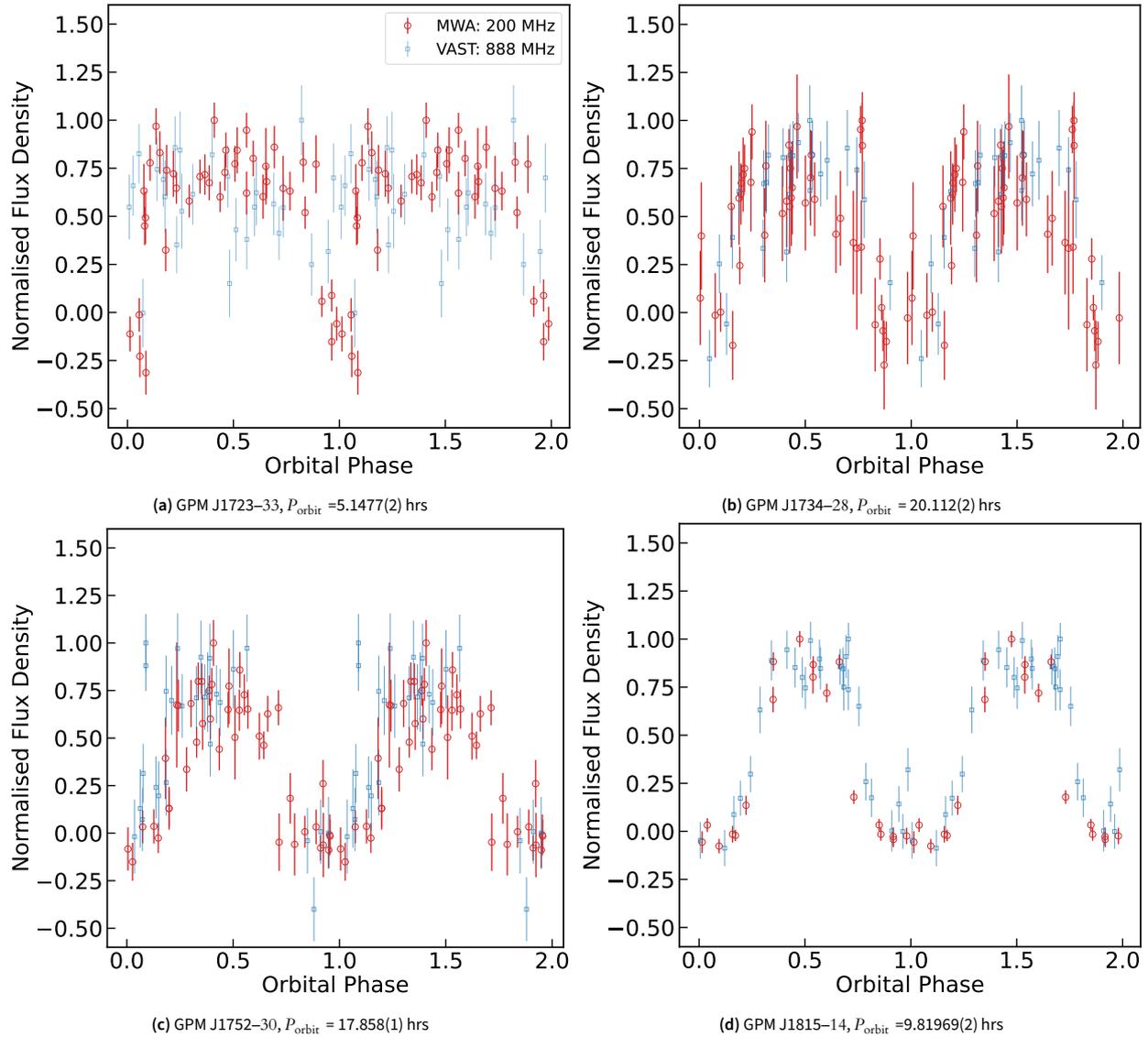

(a) GPM J1723–33, $P_{\text{orbit}}$ = 5.1477(2) hrs

(b) GPM J1734–28, $P_{\text{orbit}}$ = 20.112(2) hrs

(c) GPM J1752–30, $P_{\text{orbit}}$ = 17.858(1) hrs

(d) GPM J1815–14, $P_{\text{orbit}}$ = 9.81969(2) hrs

**Figure 6.** The normalised light curves, where the flux densities for the MWA (red points, 200 MHz) and VAST (blue points, 888 MHz) data are folded on the best-fit orbital periods of the systems. The flux densities for each dataset have been normalised to their maxima to remove the effect of the steep spectral indices of the source.

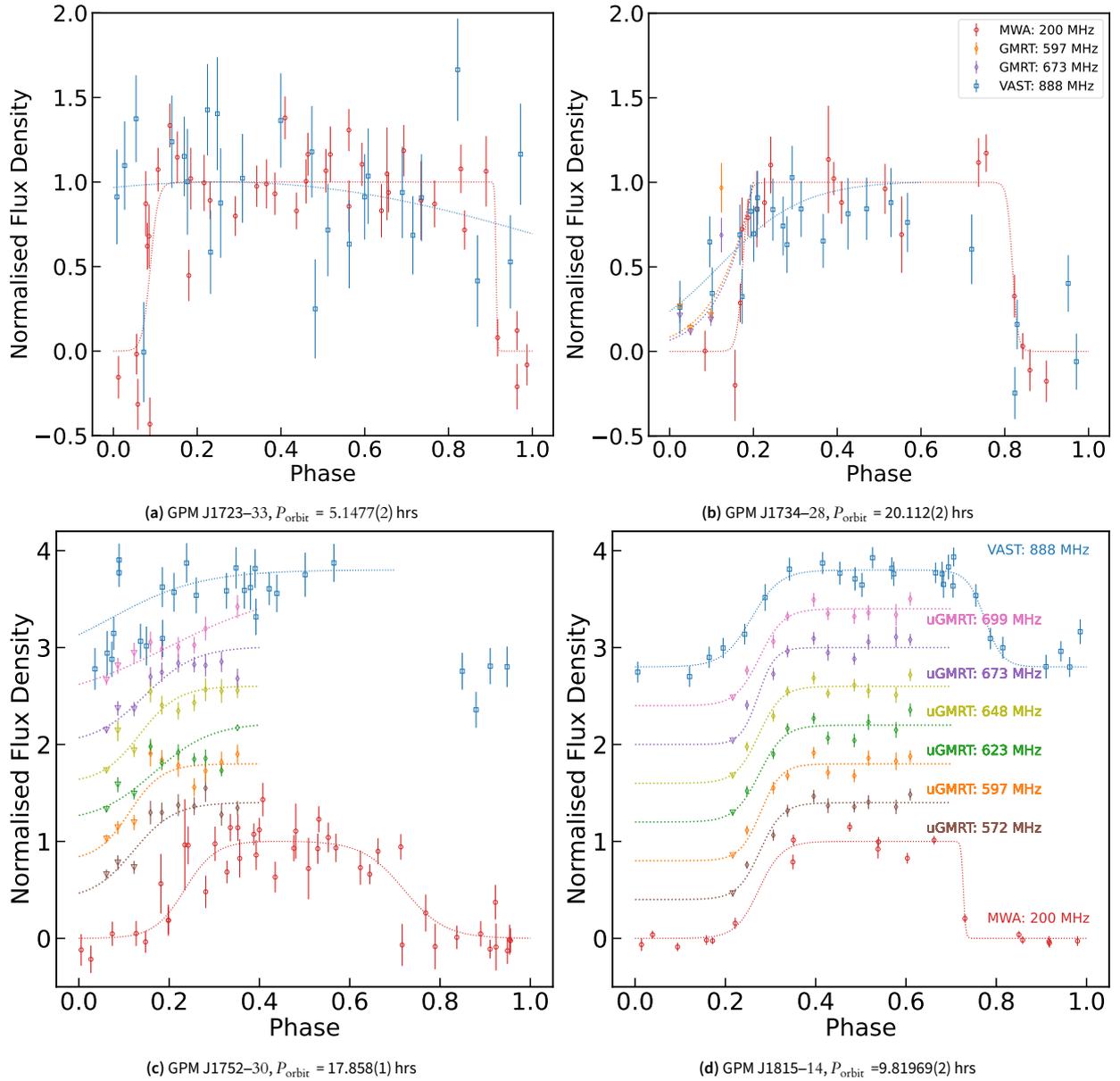

**Figure 7.** The folded light curves for each source observed with the MWA (200 MHz), uGMRT (550–750 MHz), and VAST (888 MHz), fitted with either Equation 3 or 4 at each frequency. The flux densities are normalised to the maximum flux density ($S_0$) derived from the fits. For clarity, a vertical offset of 0.5 has been added between successive frequencies relative to the MWA baseline in panels (c) and (d).

Simple power law

$$S_\nu = S_0 \left(\frac{\nu}{\nu_0}\right)^\alpha \quad (5)$$

where $S_0$ is the reference flux density at frequency $\nu_0$, and $\alpha$ is the spectral index.

Log parabola

$$S_\nu = S_0 \left(\frac{\nu}{\nu_0}\right)^{a+b\log\left(\frac{\nu}{\nu_0}\right)} \quad (6)$$

where $a$ is the spectral index and $b$ is the curvature parameter.

Broken power Law

$$S_\nu = \begin{cases} S_0 \left(\frac{\nu}{\nu_0}\right)^{a_1} & \text{for } \nu \leq \nu_b \\ S_0 \left(\frac{\nu}{\nu_0}\right)^{a_2} \left(\frac{\nu_b}{\nu_0}\right)^{a_1-a_2} & \text{for } \nu > \nu_b \end{cases} \quad (7)$$

where $a_1$ and $a_2$ are the spectral indices below and above the break frequency $\nu_b$.

The spectral models were fitted using CURVE-FIT, with uncertainties on the parameters derived from the covariance matrix of the fit. The resulting fits are shown in Figure 8 and the corresponding parameter values are given in Tabel 5. The best model was determined using Akaike information criterion (AIC) (Jankowski et al. 2018), which estimates the information lost for a given model. The best models are highlighted in the table, and all sources are found to have steep negative spectral indices.

### 4.5 Companion Modelling

We compute the range of companion mass ($m_c$) and semi-major axis (a) for a binary system using the mass function

$$f(m_p, m_c) = \frac{4\pi^2}{G}\frac{(a\sin i)^2}{P_{\text{orbit}}^2} = \frac{(m_c \sin i)^3}{(m_p + m_c)^2} \quad (8)$$

where $m_p$ is the fixed mass of the neutron star assumed to be 1.4 $M_{sun}$ (which is typically seen for pulsars (Zhang et al. 2011)), the inclination angle $i$ is fixed to $\pi/2$ (edge-on orbit), and G is the gravitational constant. From Section 3.2, we calculated an acceleration for GPM J1723–33 of 2.1493840(2) ms$^{-2}$. Using Equation (3) from Ng et al. (2015), we calculated a mass function of 5.7×10$^{-6}$ and a corresponding minimum companion mass of 0.023 $M_\odot$. Assuming the same acceleration value for the other three sources, we derived the following minimum companion masses GPM J1815–14: 0.05 $M_\odot$, GPM J1734–28: 0.15 $M_\odot$, GPM J1752–30: 0.063 $M_\odot$. These values are used as lower limits for the minimum companion mass in Equation 8.

Figure 9 plots the minimum companion mass against semi-major axis for known binary systems in the ATNF PULSAR CATALOGUE[e] (Catalogue Version 2.6.0). The range of $m_c$ values for our four systems is plotted, showing a distribution

e. https://www.atnf.csiro.au/research/pulsar/psrcat/

along the line consistent with the observed population. Assuming an upper limit of 0.4 $M_\odot$ for the companion mass, as typically observed in RB systems (Roberts 2013), we can estimate an upper limit for the semi-major axis of our systems. The upper limit for semi-major axis values calculated are as follows, GPM J1723–33: 0.410$R_\odot$, GPM J1734–28: 1.01$R_\odot$, GPM J1752–30: 0.67$R_\odot$ and GPM J1815–14: 0.63$R_\odot$.

Using these upper limits on semi-major axis, the Roche lobe radius can be computed using the Eggleton approximation (Eggleton 1983). The approximate values computed are as follows, GPM J1723–33: 0.11$R_\odot$, GPM J1734–28: 0.28$R_\odot$, GPM J1752–30: 0.19$R_\odot$ and GPM J1815–14: 0.17$R_\odot$.

As the systems are no longer accreting, we can reasonably assume that the companions are not overflowing their Roche lobes. Thus, these radii represent lower limits on the sizes of the companion stars. Assuming a lower-limit effective temperature of 3000 K, we use the Stefan–Boltzmann law to estimate the corresponding bolometric luminosities, and from there calculate lower-limit apparent magnitudes (assuming distances determined in Section 5.2). The calculated approximate lower limit on apparent magnitude are as follows, GPM J1723–33: 22, GPM J1734–28: 17, GPM J1752–30: 19, and GPM J1815–14: 19. Comparing these estimates with the DECaPS2 photometry in Table 3, we find that our lower-limit apparent magnitude estimates for GPM J1723–33 and GPM J1752–30 are consistent with sources listed in the table, specifically, those with `obj_id` 4522440858633389827 and 4954927160361131656, respectively.

### 4.6 Absorption Models

RB pulsars are seen to ablate their companion stars. This ablated material surrounds the binary system and can absorb or scatter the pulsar's radio emission, often leading to long-duration eclipses in the observed radio emission. We explore the eclipse mechanisms suggested by Thompson et al. (1994) that could be responsible for the observed eclipses. This is done using the flux density from the fitted Fermi-Dirac models (see Section 4.3) for MWA, VAST, and uGMRT. The optical depth is then computed using Equations 9 to 12, as outlined by Thompson et al. (1994).

#### 4.6.1 Cyclotron Absorption

One possible mechanism is cyclotron absorption, where the companion star and pulsar wind are in a magnetic field. Following the same approach as Thompson et al. (1994), we equate the magnetic pressure $P_B = \frac{B_E^2}{8\pi}$ of the plasma to the pulsar wind pressure $U_E = \frac{\dot{E}}{4\pi c a^2}$. Taking the spin-down power of the pulsar $\dot{E} \sim 10^{34}$ (Kansabanik et al. 2021), we can calculate the characteristic magnetic field $B_E$, the values are given in Table 10. The optical depth for cyclotron absorption is given by (Thompson et al. 1994):

$$\tau_{cyc}(\nu) = \frac{\pi}{2}\frac{m^{m+1}}{m!}\left(\frac{mk_BT}{2m_ec^2}\right)^{(m-1)}\frac{n_e e^2 L_B}{m_e c \nu} \quad (9)$$

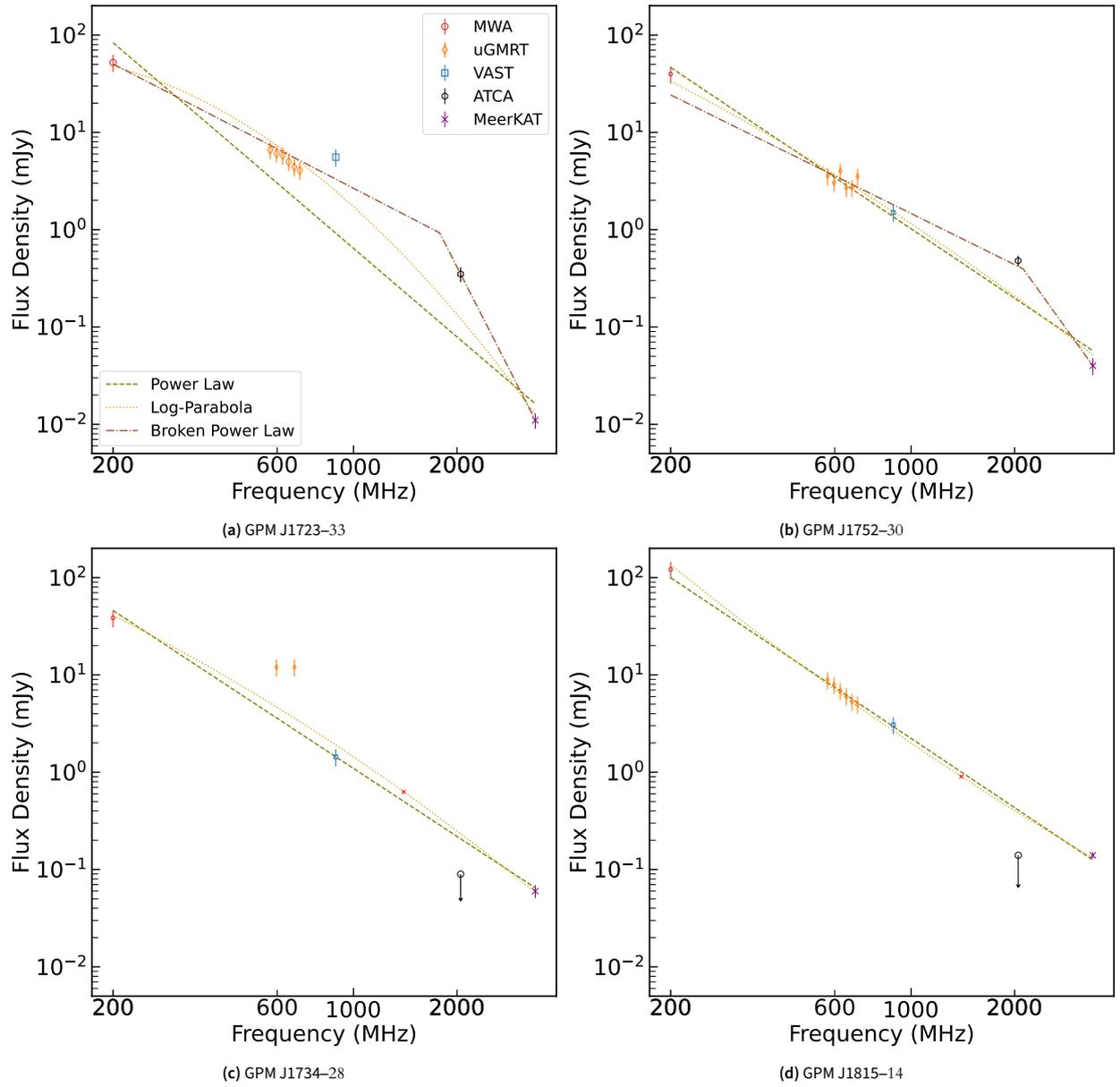

**Figure 8.** Spectra of the four sources fitted with Equations 5–7, using flux densities corresponding to the inferior conjunction of the candidate pulsars. The resulting model parameters are listed in Table 5.

**Table 5.** Details of the fitted SED parameters using Equations 5–7. We report the Akaike Information Criterion (AIC) of each model and highlight, in bold, the best-fitting model in the table, identified by the lowest AIC value.

| Source | Power Law | | Log Parabola | | | Broken Power Law | | | |
|---|---|---|---|---|---|---|---|---|---|
| | $\alpha$ | AIC | a | b | AIC | $\alpha_1$ | $\alpha_2$ | $\nu_b$ (MHz) | AIC |
| GPM J1723–33 | −3.02(6) | 109.41 | −0.70(8) | 6(1) | 32.91 | −1.8(2) | −6.9(5) | 1800(100) | **16.76** |
| GPM J1752–30 | −2.37(7) | 37.95 | −0.09(121) | −0.16(9) | 36.69 | −1.7(1) | −4.9(5) | 211.595(2) | **17.57** |
| GPM J1734–28 | −2.32(8) | 32.29 | **0.3(1.0)** | **−0.20(9)** | **30.34** | - | - | - | - |
| GPM J1815–14 | −2.36(6) | 21.97 | **−4.3(9)** | **1.37(7)** | **20.20** | - | - | - | - |

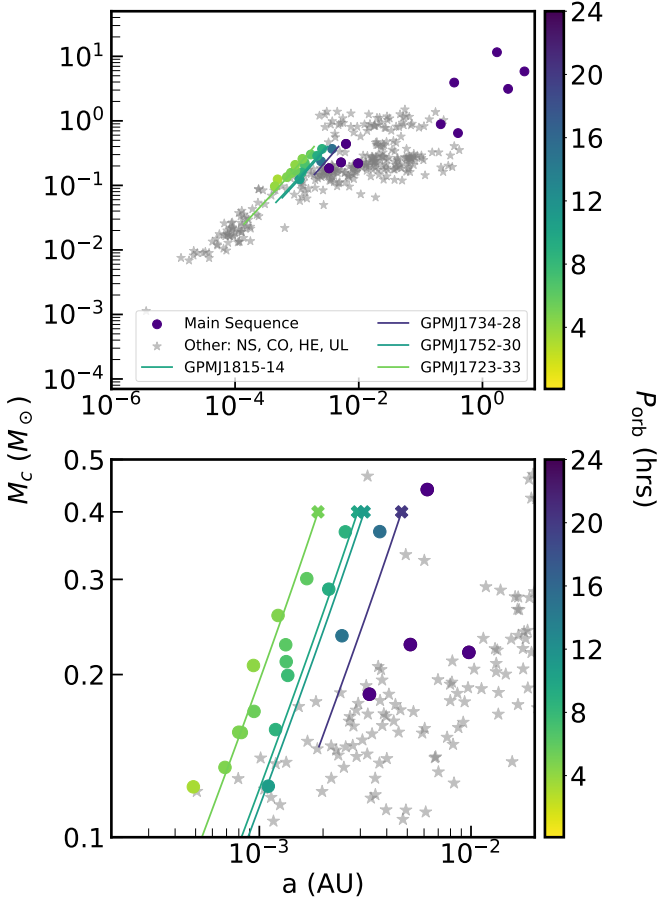

**Figure 9.** Minimum companion masses against the projected semi-major axes for known binary pulsars from the ATNF catalogue, along with the range of companion masses allowed for each of our targets, calculated using Equation 8. The colour map indicates the orbital period in hours. The bottom panel provides a zoomed-in view of the top panel. The upper limits on the minimum companion mass for our sources are marked with an 'X' along each line. Binary MSPs are categorized by companion type: Neutron Star (NS), Helium White Dwarf (He), Carbon-Oxygen White Dwarf (CO), and Ultra-light companion (UL).

where $m = \frac{\nu}{\nu_B}$ is the cyclotron harmonic, where $\nu_B = \frac{eB}{2\pi m_e c}$ is the cyclotron frequency. $T$ is the temperature in Kelvin, and $L_B = |\frac{ds}{d\ln B}|$ is the scale length of the magnetic field in cm. We assume that the scale length of electron density variations and magnetic field variations are similar for simplicity. Therefore, we take $n_e L_B$ to be equal to the average electron column density $N_e$ (Thompson et al. 1994). For Cyclotron absorption to be valid, the temperature must satisfy $T < \frac{m_e c^2}{2 k_B m^3}$.

To calculate the optical depth for our systems, we allow $N_e$ to vary between $10^4 - 10^{30} cm^{-2}$, T between $10^2 - 10^9 K$ and B between $10^{-2} - 10^2 G$, following the approach of Zic et al. (2024). The minimum temperature found to produce eclipses at 888 MHz is $> 10^8$ K, for each system.

**Table 6.** Calculated energy density $U_E$ and characteristic magnetic field $B_E$ for cyclotron absorption, see Section 4.6.1.

| Source | $\sim U_E$ | $\sim B_E$ |
|---|---|---|
|  | erg cm$^{-3}$ | G |
| GPM J1723–33 | 4.56 | 11 |
| GPM J1752–30 | 6.23 | 13 |
| GPM J1734–28 | 12.1 | 17 |
| GPM J1815–14 | 14.0 | 19 |

#### 4.6.2 Compton Scattering

For Compton scattering, the optical depth is given as (Thompson et al. 1994):

$$\tau_{ind}(\nu) = 4 \times 10^6 \frac{N_e S_\nu}{\nu^2} \langle f(\phi) \rangle |\alpha + 1| \cdot \left(\frac{d_{kpc}}{a}\right)^2 M \quad (10)$$

where $S_\nu$ is the mean flux density of the pulsar in mJy at frequency $\nu$ MHz, $\alpha$ is the spectral index of the incident radiation, $d_{kpc}$ is the distance to the scattering centre from the observer in kiloparsecs and $a$ is the distance between the companion and the pulsar in cm. $\langle f(\phi) \rangle$ is the angular factor which is averaged out over the scattering region and M is the demagnefication factor (Kumari et al. 2024) which can be between 0 and 1. The distance of GPM J1723–33 is $\sim$9.7 kpc obtained using the YMW16 model, and we have assumed a distance of 8 kpc for the other three sources (see Section 3.2 for details). The maximum $\tau_{ind}$ found for each system is much less than 1.

#### 4.6.3 Free-Free Absorption

For free-free absorption, the optical depth is given by (Thompson et al. 1994):

$$\tau_{ff}(\nu) \approx 3.1 \times 10^{-8} \cdot \frac{f_{cl} N_e^2}{T^{3/2} \nu^2 L} \cdot \ln\left(5 \times 10^{10} \cdot \frac{T^{3/2}}{\nu}\right) \quad (11)$$

where $f_{cl}$ is the clumping factor, $T$ is the temperature, and $L$ is the absorption length. We fix the absorption length as $L = 2R_E$, where the eclipse radius is defined as $R_E = \pi a \Delta \phi$. Where, $a$ is the semi-major axis calculated in Section 4.5, and $\Delta \phi = \phi_e - \phi_i$ is the eclipse duration orbital phase.

To constrain the physical parameters for free-free absorption in Equation 11, we performed a Markov Chain Monte Carlo (MCMC) analysis. The details of the MCMC setup are provided in Section 4.6.5.

#### 4.6.4 Synchrotron Absorption

Synchrotron absorption arises from a population of non-thermal electrons, often assumed to possess a power-law energy distribution $n(E) = n_0 E^{-p}$. The optical depth is given by (Thompson et al. 1994):

$$\tau_{sync}(\nu) = \left(\frac{3^{\frac{(p+1)}{2}} \Gamma\left(\frac{3p+2}{12}\right) \Gamma\left(\frac{3p+22}{12}\right)}{4}\right) \left(\frac{\sin\theta}{m}\right)^{\frac{p+2}{2}} \frac{n_0 e^2}{m_e c \nu} L \quad (12)$$

Where $L$ is the absorption length, $p$ is the power-law index, $\theta$ is the angle of the magnetic field lines with our line of sight, $n_0$ is the non-thermal electron density, $m_e$ is the mass of the electron, $m$ is the cyclotron harmonic at frequency $\nu$, $e$ is the charge on the electron and $c$ is the speed of light. We fix the absorption length as $L = 2R_E$ (as in Equation 11), and a fixed viewing angle of $\pi/3$. An MCMC analysis is performed to constrain the physical parameters in Equation 12, with further details presented in Section 4.6.5.

### 4.6.5 MCMC

We perform an MCMC analysis to constrain the physical parameters for free-free and synchrotron absorption, using the `emcee` sampler. We modelled the observed eclipse durations as a function of frequency using the models given in Equation 11 and 12. Our MCMC setup and parameter constraints follow those used in Zic et al. (2024).

Sampling was performed in logarithmic space, using uniform priors over the following ranges: for free-free absorption, $\log(f_{cl}) \in (0, 10)$, $\log(T) \in (0, 9)$ and $\log(N_e) \in (10, 20)$; for synchrotron absorption, $\log(B) \in (-2, 2)$, $\log(p) \in (0, 8)$ and $\log(n_0) \in (0, 8)$.

The log-likelihood function compares how well the modelled optical depths, $\tau_{ff(\nu)}$ and $\tau_{sync(\nu)}$, fits the observed optical depth measurements $\tau(\nu)$, assuming gaussian errors $\tau_{err(\nu)}$.

We calculated $\tau(\nu)$ at frequencies corresponding to MWA (200 MHz), uGMRT (500–600 MHz) and VAST (888 MHz) at the phase just after the eclipse. This was done using the radiative transfer equation $S = S_0(\nu)e^{-\tau(\nu)}$, where $S_0(\nu)$ is the flux density at the pulsars' inferior conjunctions (from Equation 3).

The log-likelihood is then given by

$$\log L = -\frac{1}{2} \sum \left( \frac{\tau(\nu) - (\tau_{ff}(\nu)/\tau_{sync}(\nu))}{\tau_{err}(\nu)} \right)^2 \quad (13)$$

and the total log-probability is the sum of the log-prior and the log-likelihood. Figure 12 and 13 present the marginal posterior distributions for the free-free and synchrotron absorption parameters, the results are discussed in Section 5.3.

### 4.7 Fermi gamma-ray properties

Section 3.1 revealed possible *Fermi* γ-ray associations for each source. The two primary requirements for γ-ray emission from pulsar candidates are: they have a spectral curvature significance higher than 3σ and the variability index is lower than 27.7 (Abdollahi et al. 2020). Pulsars are expected to be steady γ-ray emitters, hence, their variability index should be low. The spectral curvature significance quantifies how much the γ-ray spectrum deviates from a simple power law, with pulsars expected to have curved spectra. Figure 10 plots the γ-ray spectral curvature significance against the γ-ray variability index (taken from the 4FGL catalogue), for γ-ray binary pulsars taken from the ATNF pulsar catalogue (Manchester et al. 2005) along with the four sources in this paper. All four of our systems align with the characteristics of the currently known binary pulsar population.

We compiled γ-ray fluxes at 100 MeV ($E_{100}$) and radio flux densities at 1400 MHz ($S_{1400}$) for known γ-ray binary pulsars in the ATNF catalogue. The $E_{100}$ values are taken from the 4FGL catalogue (Abdollahi et al. 2022) and $S_{1400}$ from the ATNF pulsar catalogue (Manchester et al. 2005). In Figure 10 we show $S_{1400}$ as a function of $E_{100}$ for all γ-ray binary pulsars, together with our four systems. There is no evident correlation between radio and γ-ray flux. All four of our systems lie within the observed range of both $E_{100}$ and $S_{1400}$ values for the current population of binary pulsars.

## 5. Discussion

### 5.1 Nature of the sources

We have discovered four radio sources in the image domain, all exhibiting eclipse-like variability, with orbital periods <1 day. Their steep negative spectral indices and potential *Fermi* associations are consistent with MSPs. One of the sources, GPM J1723–33, has been confirmed as a MSP through pulsation detections. In Section 4.3, all systems were found to have eclipse durations lasting between ~30–50 % of their orbital phase. The only known class of objects that can explain these sources would be "spider" systems, where the eclipses are due to the intra-binary material. Our sources are more consistent with the properties of RB systems than BW, as they exhibit longer duration eclipses and are found to have main-sequence companions consistent with RB systems.

### 5.2 Non-detection of radio pulsations

Our search for radio pulsations made an 8σ detection of GPM J1723–33. One possible reason for the non-detection of the remaining three sources is their intrinsic faintness. However, since simultaneous imaging and beamforming observations were taken with uGMRT and MeerKAT, and all sources were detected with sufficient S/N in images (ranging from approximately 4.4–18), they should be intrinsically bright enough to be detectable in the time domain. For the Parkes observations, we estimate the minimum detectable pulsed flux density using the radiometer equation (Lorimer and Kramer 2004). Assuming a 15 minute integration, a canonical MSP duty cycle of 10%, a gain of 1.8 J/K, a system temperature of 21 K, and a bandwidth of 1024 MHz, we calculate a minimum detectable phase-averaged flux density of 0.012 mJy for a S/N of 10. Given that GPM J1734–28 has the smallest measured flux density of 0.14 mJy at 2368 MHz of our sources, all sources should be well within Parkes' detection capability.

Another plausible explanation for the non-detections of the other sources is the presence of ISM scattering or eclipsing material. Since these systems are apparently close to the Galactic bulge, an area heavily impacted by ISM scattering, detecting pulsars via their pulsations in this region can be challenging, particularly at lower frequencies (e.g., uGMRT Band-4), because the scattering timescale scales with $\nu^{-4}$.

Assuming GPM J1734–28, GPM J1752–30 and GPM J1815–14 are indeed MSPs and the non-detections of pulses are solely due to scattering in the ISM, we can estimate a lower limit on their distance based on the electron density models of the region. To explain the lack of pulsations, we assume a lower limit of 1 ms for $\tau_{sc}$ at 3 GHz. Using the YMW16 model (Yao, Manchester, and Wang 2017), the distances required to obtain this scattering timescale are found to be greater than 25 kpc. Therefore, GPM J1734–28, GPM J1752–30 and GPM J1815–14 would lie outside the Galactic Disk. These results indicate that the observed scattering of pulsations is most likely caused by intra-binary material.

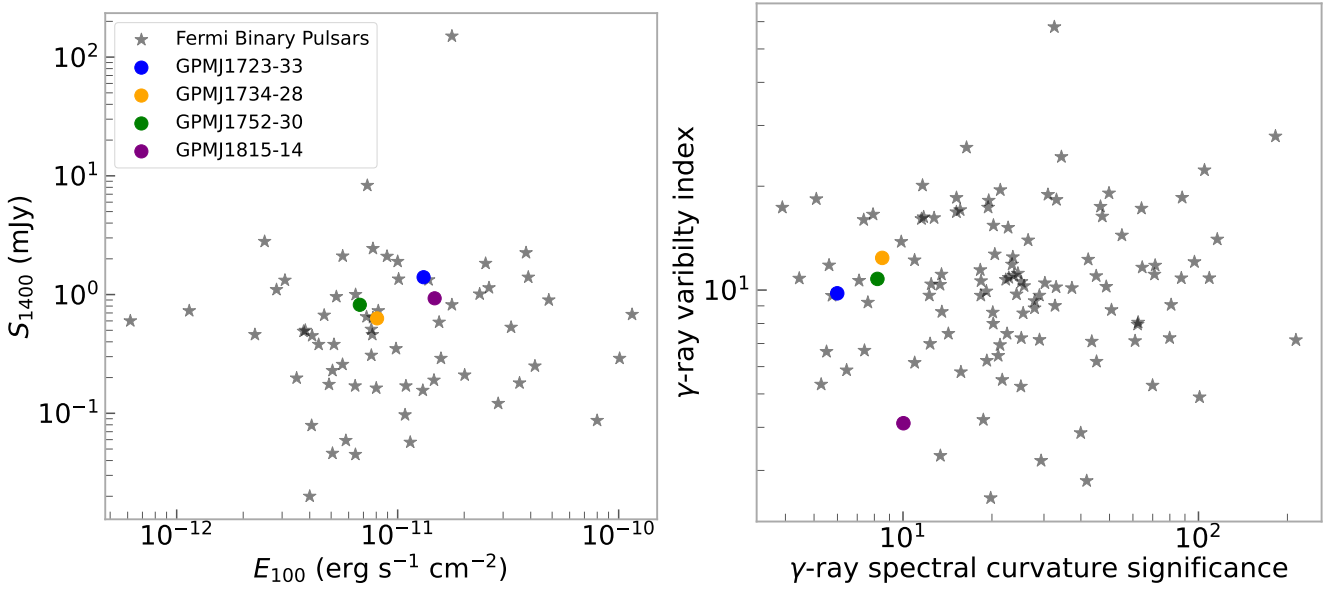

**Figure 10.** *left:* Radio flux density at 1400 MHz ($S_{1400}$) plotted against *Fermi* γ-ray energy flux ($E_{100}$) at 100 MeV. There is no evident correlation between radio and γ-ray flux. All four of our systems lie within the observed range of both $E_{100}$ and $S_{1400}$ values for the current population of binary pulsars. *right:* γ-ray variability index against spectral curvature significance. The plots display the four sources from this paper alongside ATNF binary pulsars with *Fermi* associations (see Section 4.7).

Figure 11 plots the locations of ATNF binary MSPs and MSPs, along with our sources. The distance of GPM J1723–33, estimated using the YMW16 model, is ∼9.7 kpc and is plotted accordingly. The other three sources are assumed to be at a distance of 8 kpc, where a large number of MSPs are expected to be located (Gonthier et al. 2018). The plot highlights the small number of known binary MSPs in the direction of the Galactic Bulge, where our systems would reside under this distance assumption.

### 5.3 Eclipse Mechanisms

In the following section, we discuss the results obtained in Section 4.6 to see what eclipse mechanism could be ruled out.

For cyclotron absorption to be valid according to Equation 9 in the non–relativistic regime, the approximation holds for temperatures $T \lesssim \left(\frac{m_e c^2}{2km^3}\right)$ (Thompson et al. 1994). We find that the temperature to produce eclipses are $> 10^8$ K which exceeds the threshold for cyclotron absorption. For Induced Compton scattering (see Section 4.6.2) the optical depth values calculated using Equation 10 are all $\ll 1$. Therefore, Induced Compton Scattering is unlikely to cause the observed eclipses.

Figure 12 gives the marginal posterior probability densities on each of the varied parameters for free-free absorption: $f_{cl}$, $T_e$, and $N_e$ for our four sources. The posterior distributions indicate that the parameters are not well-constrained across all sources. Specifically, $N_e$ and $f_{cl}$ are pushed against the upper bounds of their prior ranges, suggesting a preference for higher values within the allowed range. This lack of constraint limits our ability to precisely characterize the physical conditions of the absorbing plasma, meaning we cannot rule out free-free absorption as a possible mechanism.

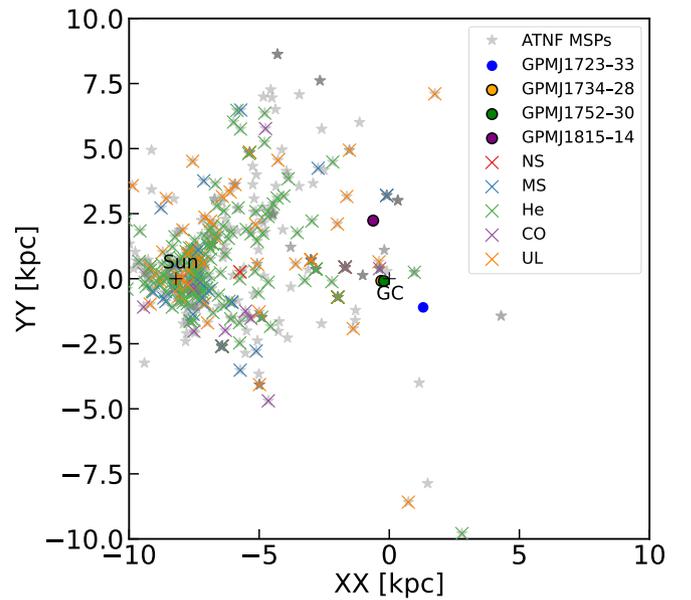

**Figure 11.** The figure gives the distances of the ATNF MSPs and binary MSPs from the Galactic Center in kpc, alongside our sources. GPM J1723–33 is plotted at a distance of 9.7 kpc determined using the YMW16 model (see Section 5.2), while the other three sources, shown with a black outline, have been plotted at an assumed distance of 8 kpc. Binary MSPs are categorized by companion type: Neutron Star (NS), Main Sequence (MS), Helium White Dwarf (He), and Ultra-light companion (UL).

Figure 13 shows the marginal posterior probability densities for the varied parameters for synchrotron absorption: $p$, $B$, and $n_0$. The posteriors indicate that the parameters are not very well constrained, with some parameter distributions returning the priors, suggesting the data provides insufficient information to meaningfully narrow the parameter space. Although better solutions may exist in an expanded parameter space, they would fall outside physically plausible bounds and are therefore not considered viable. Given these limitations, synchrotron absorption remains a viable eclipse mechanism and cannot be ruled out with the current observations.

We conclude that the models are not sufficiently constrained to rule out either free-free or synchrotron absorption, and both remain plausible eclipse mechanisms. Additional observational constraints on the plasma density, such as measurements of dispersion measure (DM) variations near eclipse, would significantly improve our understanding of the underlying eclipse mechanism.

### 5.4 Pseudo-Luminosity distribution of RB pulsars

In the following section, we calculate the pseudo-luminosities of the four RB candidates discussed in this paper. We then explore the luminosity distribution of known RB pulsars, adding GPM J1723–33 to the population, as this is the only source with detected pulsations and thus a distance measurement. By fitting a model to the observed luminosity distribution of RB pulsars, we assess where the luminosities of the remaining sources lie relative to the known population. Finally, we estimate how many additional RBs could be discovered in the future due to the development of increased sensitivity instruments such as SKA–low.

Traditionally, luminosity has been considered as a function of the spin parameters of pulsars (Bagchi 2013). Parameter-independent models, such as log-normal and power-law distributions, have also been considered to fit the observed luminosities. Faucher-Giguère and Kaspi 2006 first established that a lognormal model provides a superior fit to the observed luminosities compared to the power law. There have been many studies fitting this model to the luminosity distribution of pulsars (e.g. Berteaud et al. 2024; Chennamangalam et al. 2013; Bagchi, Lorimer, and Chennamangalam 2011). There haven't been any spider pulsar-specific studies.

Using the distance found for GPM J1723–33 (defined in Section 5.2) and an assumed distance of 8 kpc for the other three sources, along with the $S_{200}$ values in Table 1 (scaled to $S_{1400}$ using the spectral indices in Table 5), we can calculate the pseudo-luminosities ($L_{1400}$) of the four sources presented in this paper. Pseudo-luminosity is defined as

$$L_\nu = S_\nu d^2 \quad (14)$$

which is expressed in units of mJy kpc$^2$. The values for $L_{1400}$ are as follows, GPM J1723–33: 130 mJy kpc$^2$, GPM J1734–28: 40 mJy kpc$^2$, GPM J1752–30: 52 mJy kpc$^2$, and GPM J1815–14: 60 mJy kpc$^2$.

We assume a lognormal distribution, as suggested by previous studies (Berteaud et al. 2024; Chennamangalam et al. 2013; Bagchi, Lorimer, and Chennamangalam 2011), of the following form:

$$f(\log(L_\nu)) = \frac{1}{\sqrt{2\pi}\sigma} \exp\left(-\frac{1}{2}\left(\frac{\log(L_\nu) - \mu}{\sigma}\right)^2\right) \quad (15)$$

where $\mu$ and $\sigma$ are the mean and width of the luminosity function.

The number of pulsars with luminosities greater than or equal to $L_\nu$ can be expressed using the complementary cumulative frequency distribution function (CCFDF), given as

$$N(\geq L_\nu) = N_{\text{tot}}\left[1 - \Phi\left(\frac{\log L_\nu - \mu}{\sigma}\right)\right] \quad (16)$$

where $\Phi$ is the cumulative distribution function.

Figure 14 shows the CCFDF at 1400 MHz for the 13 RB pulsars (i.e. those with main-sequence companions) in the ATNF catalogue with luminosity and flux density values at 1400 MHz, along with GPM J1723–33. By fitting Equation 16 to the combined sample, we find best-fit parameters of $\mu = 2.4 \pm 0.6$ and $\sigma = 2.0 \pm 0.3$, where the uncertainties are estimated using bootstrapping (Imbens and Menzel 2018), with 1000 resamples of the luminosity data.

GPM J1723–33 lies at the more luminous end of the known population; assuming a distance of 8 kpc, the other sources are similar but slightly less luminous. To illustrate the detectability range of our search, we include luminosity limits at a distance of 8 kpc (where a large population of undiscovered MSPs could lie (Gonthier et al. 2018)) for both the MWA and the future SKA–low telescope, scaled to 1400 MHz.

Within the luminosity range between the SKA and MWA detection thresholds, i.e., $L_{\text{SKA}} \leq L \leq L_{\text{MWA}}$, we find $N = 11$ pulsars. These results imply that SKA–low has the sensitivity to detect roughly three times the number of sources identified in the GPM with the MWA, due to its significantly higher sensitivity.

This highlights the potential of next-generation low-frequency radio telescopes, such as SKA–Low, to detect a larger population of spider pulsars in the image domain. Other factors would need to be taken into account to detect spider pulsars via their eclipses, including their inclination angle and the density of the eclipsing material. Targeted searches for spider pulsars in the locations of unassociated *Fermi* sources using instruments like SKA–Low could confirm whether a significant fraction of these sources are indeed MSPs. As well as the increased sensitivity, the full SKA-Low field of view will cover most Fermi error ellipses, making it ideal for these searches.

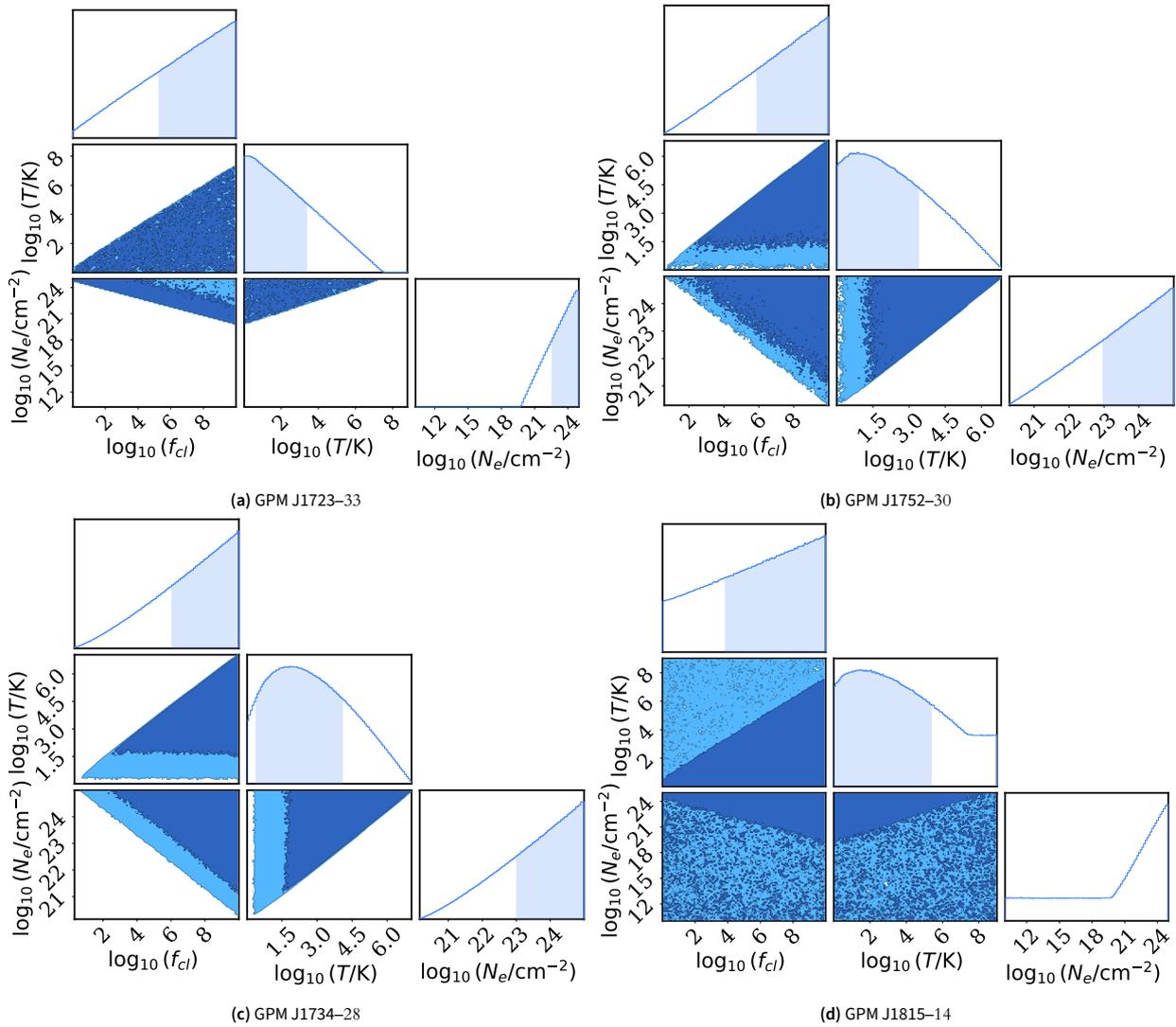

**Figure 12.** Posterior distributions for free-free absorption, where $f_{cl}$ is the clumping factor, $T$ is the temperature and $N_e$ is the electron column density. See Section 4.6.5 for more details.

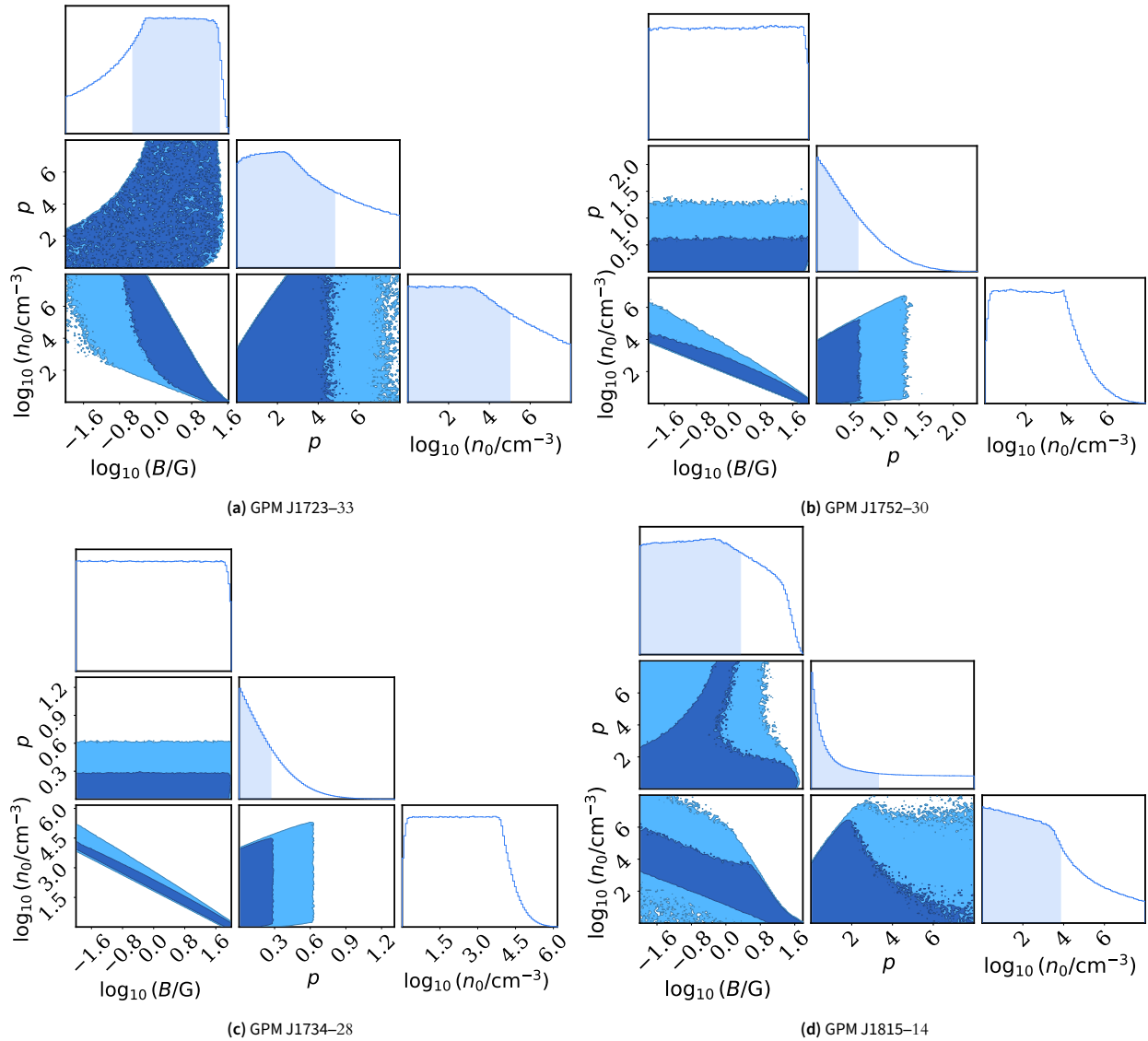

**Figure 13.** Posterior distributions for synchrotron absorption, where $B$ is the magnetic field strength, $n_0$ is the non-thermal electron density and $P$ is the power-law index. See Section 4.6.5 for more details.

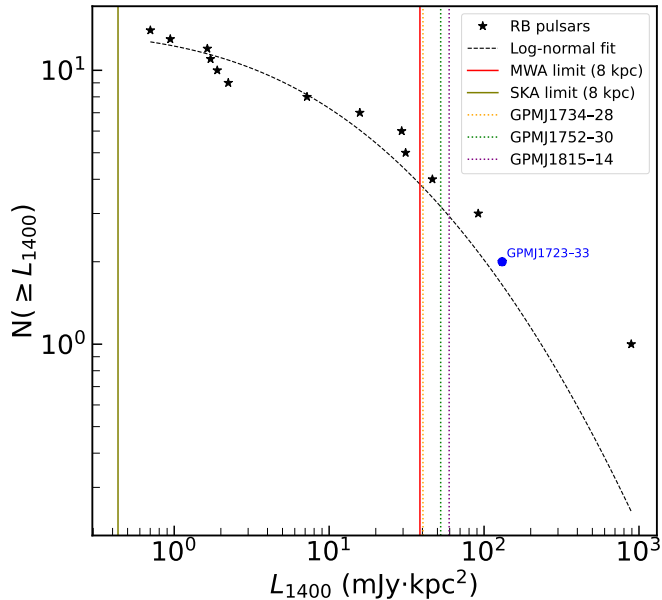

**Figure 14.** CCFDF of luminosities at 1400 MHz for RB pulsars in the ATNF catalogue, along GPM J1723–33 (at a distance of 9.7 kpc) included in the population. The distribution is fitted with a log-normal model, as described in Equation 16. Vertical lines indicate the sensitivity limits of SKA–low and MWA, as well as the pseudo-luminosities of the other three sources discussed in this paper, which are assumed to sit at a distance of 8 kpc.

## 6. Summary

We have presented the discovery and follow-up of four candidate RB spider pulsars: GPM J1723–33, GPM J1734–28, GPM J1752–30 and GPM J1815–14. These objets have potential *Fermi* γ-ray associations, with GPM J1723–33 and GPM J1815–14 lying within a *Fermi* 95 % error ellipse. All four sources exhibit radio variability consistent with eclipses and are found to have short periods (< 1 day), aligning with the typical characteristics of spider pulsars.

Using MeerKAT, we searched for pulsations and obtained a detection of GPM J1723–33, while no pulsations have been confirmed from the other three sources. This suggests that the pulses are scattered at low frequencies (below 3.5 GHz), prompting the need for searches at higher frequencies. However, the systems are thought to lie near the Galactic bulge, they may be scattered beyond the reach of our detectability with current instruments. As discussed in Section 3.4, we identified potential optical counterparts to GPM J1723–33, GPM J1734–28 and GPM J1752–30, but all four fields are strongly affected by dust extinction, and the true counterparts may be obscured. Inspection of near-IR imaging of the fields did not reveal any counterparts, but the available survey imaging uses very short exposures. Follow-up with larger optical telescopes and longer near-IR exposures is needed to identify the true counterparts. Some initial constraints on the companion properties were made in Section 4.5. By analysing the eclipse properties from our fitted Fermi-Dirac models (see Section 4.3), we explored potential eclipse mechanisms and ruled out Compton scattering and cyclotron absorption. Additionally, long-term observations with the MWA revealed that the systems are highly dynamic, altering the eclipse durations over time. Therefore, a single eclipse mechanism may not be sufficient to explain all observed eclipses.

As shown in Section 5.4, our confirmed pulsar lies towards the high luminosity end of the distribution of known RB pulsars. By fitting a model to this distribution, we estimate that SKA–Low could detect approximately three times as many RB pulsars using image-plane searches, owing to its significantly improved sensitivity.

The identification of potential radio counterparts to four previously unassociated *Fermi* sources brings us one step closer to understanding how many *Fermi* unassociated sources could be MSPs, and ultimately to uncovering the origin of the γ-ray excess. The development of sensitive, wide field-of-view telescopes such as SKA–Low will greatly enhance the effectiveness of such searches. In future work, we aim to perform a population analysis to estimate how many spider pulsars could account for currently unassociated *Fermi* sources.

**Acknowledgments** N.H.-W. is the recipient of an Australian Research Council Future Fellowship (project number FT190100231). R.K. acknowledges support from NSF grant AST-2205550.

This scientific work uses data obtained from Inyarrimanha Ilgari Bundara, the CSIRO Murchison Radio-astronomy Observatory. Support for the operation of the MWA is provided by the Australian Government (NCRIS), under a contract to Curtin University administered by Astronomy Australia Limited. ASVO has received funding from the Australian Commonwealth Government through the National eResearch Collaboration Tools and Resources (NeCTAR) Project, the Australian National Data Service (ANDS), and the National Collaborative Research Infrastructure Strategy. The Australian SKA Pathfinder is part of the Australia Telescope National Facility which is managed by CSIRO. Operation of ASKAP is funded by the Australian Government with support from the National Collaborative Research Infrastructure Strategy. ASKAP and the MWA use the resources of the Pawsey Supercomputing Centre. Establishment of ASKAP, Inyarrimanha Ilgari Bundara, and the Pawsey Supercomputing Centre are initiatives of the Australian Government, with support from the Government of Western Australia and the Science and Industry Endowment Fund. We acknowledge the Wajarri Yamaji People as the Traditional Owners and Native Title Holders of the observatory site.

The MeerKAT telescope is operated by the South African Radio Astronomy Observatory, which is a facility of the National Research Foundation, an agency of the Department of Science and Innovation. This work has made use of the "MPIfR S-band receiver system" designed, constructed and maintained by funding of the MPI für Radioastronomy and the Max-Planck-Society. Observations made use of the Pulsar Timing User Supplied Equipment (PTUSE) servers at MeerKAT which were funded by the MeerTime Collaboration members ASTRON, AUT, CSIRO, ICRAR-Curtin, MPIfR, INAF, NRAO, Swinburne University of Technology, the University of Oxford, UBC and the University of Manch-ester. The system design and integration was led by Swinburne University of Technology and Auckland University of Technology in collaboration with SARAO and supported by the ARC Centre of Excellence for Gravitational Wave Discovery (OzGrav) under grant CE170100004.

The Australia Telescope Compact Array is part of the Australia Telescope National Facility (https://ror.org/05qajvd42) which is funded by the Australian Government for operation as a National Facility managed by CSIRO. We acknowledge the Gomeroi people as the Traditional Owners of the Observatory site.

Murriyang, CSIRO's Parkes radio telescope, is part of the Australia Telescope National Facility (https://ror.org/05qajvd42) which is funded by the Australian Government for operation as a National Facility managed by CSIRO. We acknowledge the Wiradjuri people as the Traditional Owners of the Observatory site.

We thank the staff of the GMRT that made these observations possible. GMRT is run by the National Centre for Radio Astrophysics of the Tata Institute of Fundamental Research.

This project used data obtained from the Dark Energy Camera (DECam), which was constructed by the Dark Energy (DES) collaboration. Funding for the DES Projects has been provided by the U.S. Department of Energy, the U.S. National Science Foundation, the Ministry of Science and Education of Spain, the Science and Technology Facilities Council of the United Kingdom, the Higher Education Funding Council for England, the National Center for Supercomputing Applications at the University of Illinois at Urbana-Champaign, the Kavli Institute of Cosmological Physics at the University of Chicago, the Center for Cosmology and Astro-Particle Physics at the Ohio State University, the Mitchell Institute for Fundamental Physics and Astronomy at Texas A&M University, Financiadora de Estudos e Projetos, Fundação Carlos Chagas Filho de Amparo à Pesquisa do Estado do Rio de Janeiro, Conselho Nacional de Desenvolvimento Científico e Tecnológico and the Ministério da Ciência, Tecnologia e Inovação, the Deutsche Forschungsgemeinschaft, and the Collaborating Institutions in the Dark Energy Survey. The Collaborating Institutions are Argonne National Laboratory, the University of California at Santa Cruz, the University of Cambridge, Centro de Investigaciones Energéticas, Medioambientales y Tecnológicas-Madrid, the University of Chicago, University College London, the DES-Brazil Consortium, the University of Edinburgh, the Eidgenössische Technische Hochschule (ETH) Zürich, Fermi National Accelerator Laboratory, the University of Illinois at Urbana-Champaign, the Institut de Ciències de l'Espai (IEEC/CSIC), the Institut de Física d'Altes Energies, Lawrence Berkeley National Laboratory, the Ludwig-Maximilians Universität München and the associated Excellence Cluster Universe, the University of Michigan, the National Optical Astronomy Observatory, the University of Nottingham, the Ohio State University, the OzDES Membership Consortium, the University of Pennsylvania, the University of Portsmouth, SLAC National Accelerator Laboratory, Stanford University, the University of Sussex, and Texas A&M University.

The Pan-STARRS1 Surveys (PS1) and the PS1 public science archive have been made possible through contributions by the Institute for Astronomy, the University of Hawaii, the Pan-STARRS Project Office, the Max-Planck Society and its participating institutes, the Max Planck Institute for Astronomy, Heidelberg and the Max Planck Institute for Extraterrestrial Physics, Garching, The Johns Hopkins University, Durham University, the University of Edinburgh, the Queen's University Belfast, the Harvard-Smithsonian Center for Astrophysics, the Las Cumbres Observatory Global Telescope Network Incorporated, the National Central University of Taiwan, the Space Telescope Science Institute, the National Aeronautics and Space Administration under Grant No. NNX08AR22G issued through the Planetary Science Division of the NASA Science Mission Directorate, the National Science Foundation Grant No. AST-1238877, the University of Maryland, Eotvos Lorand University (ELTE), the Los Alamos National Laboratory, and the Gordon and Betty Moore Foundation.

The VISTA Data Flow System pipeline processing and science archive are described in Irwin et al. 2004, Hambly et al. 2008 and Cross et al. 2012.

Basic research in radio astronomy at NRL is supported by the Chief of Naval Research (CNR). Parts of this research were supported by the Australian Research Council Centre of Excellence for Gravitational Wave Discovery (OzGrav), project number CE230100016. I want to thank Drs. Scott Hyman, Namir Kassim, and Andy Wang for their contributions to proposals related to this project. I also thank Dr. Simona Giacintucci, Apurba Bera, Kat Ross, and Adelle Goodwin for their expertise in data reduction related to this project. Additionally, I would like to acknowledge Dr. Tim Galvin for his assistance in writing the processing pipeline for the Galactic Plane Monitor.

# Appendix 1. Lafler–Kinman's String Length (LKSL) Periodograms

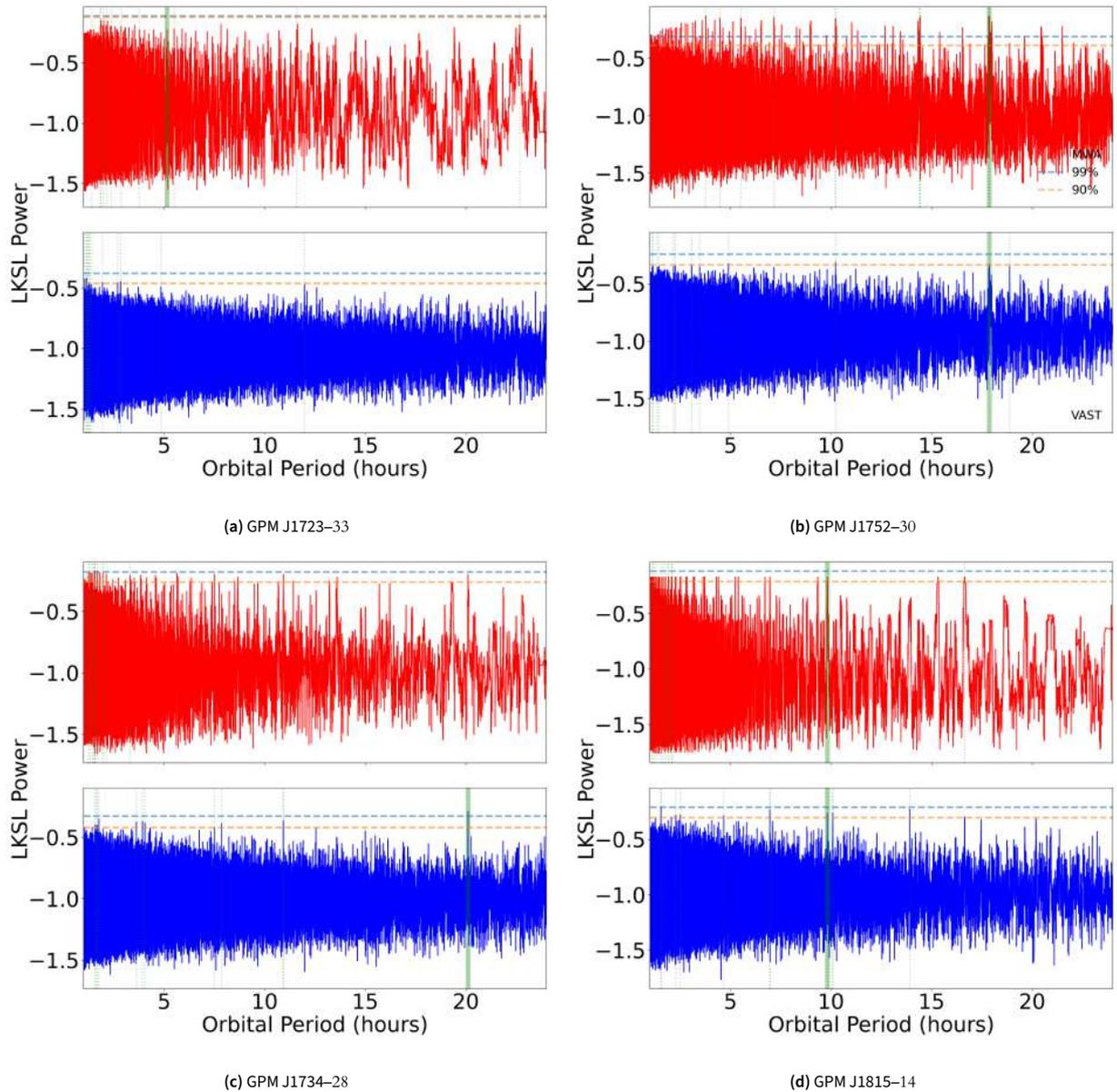

(a) GPM J1723−33

(b) GPM J1752−30

(c) GPM J1734−28

(d) GPM J1815−14

**Figure 15.** Figure shows the LKSL periodograms for the MWA and VAST data sets for each source. The 90 and 99 % confidence thresholds for periodicity, calculated using bootstrap methods, are highlighted on the plots. The best orbital period (see Section 4.2 for how this is selected) is highlighted in green.